\documentclass[usenatbib,usedcolumn,usegraphicx]{mn2e}

\title[The mass-metallicity relation at $z\sim1.4$]{The mass-metallicity relation at $z\sim1.4$ revealed with Subaru/FMOS\thanks{Based on data collected at Subaru Telescope, which is operated by the National Astronomical Observatory of Japan.}}
\author[Yabe et al.]{Kiyoto Yabe$^{1,2}$\thanks{E-mail: kiyoto.yabe@nao.ac.jp}, Kouji Ohta$^{2}$\thanks{E-mail: ohta@kusastro.kyoto-u.ac.jp}, Fumihide Iwamuro$^{2}$, Masayuki Akiyama$^{3}$,
\newauthor
Naoyuki Tamura$^{4,5}$, Suraphong Yuma$^{2, 6}$, Masahiko Kimura$^{4,7}$, Naruhisa Takato$^{4}$, 
\newauthor
Yuki Moritani$^{2,8}$, Masanao Sumiyoshi$^{2}$, Toshinori Maihara$^{2}$, John Silverman$^{5}$, 
\newauthor
Gavin Dalton$^{9,10}$, Ian Lewis$^{9}$, David Bonfield$^{11}$, Hanshin Lee$^{12}$, Emma Curtis-Lake$^{13}$, 
\newauthor
Edward Macaulay$^{9}$, and Fraser Clarke$^{9}$\\
$^{1}$ Division of Optical and Infrared Astronomy, National Astronomical Observatory of Japan, 2-21-1, Osawa, Mitaka, 181-8588, Japan\\
$^{2}$ Department of Astronomy, Kyoto University, Sakyo-ku, Kyoto, 606-8502, Japan\\
$^{3}$ Astronomical Institute, Tohoku University, Aoba-ku, Sendai, 980-8578, Japan\\
$^{4}$ Subaru Telescope, National Astronomical Observatory of Japan, 650 North A'ohoku Place, Hilo, HI 96720, USA\\
$^{5}$ Kavli Institute for the Physics and Mathematics of the Universe, The University of Tokyo, Kashiwanoha, Kashiwa, 277-8583, Japan\\
$^{6}$ Institute for Cosmic Ray Research, The University of Tokyo, 5-1-5 Kashiwanoha, Kashiwa, 277-8582, Japan\\
$^{7}$ Institute of Astronomy and Astrophysics, Academia Sinica, P. O. Box 23-141, Taipei 10617, Taiwan\\
$^{8}$ Hiroshima Astrophysical Science Center, Hiroshima University, 1-3-1 Kagamiyama, Higashi-Hiroshima, 739-8526, Japan\\
$^{9}$ Department of Astrophysics, University of Oxford, Keble Road, Oxford OX1 3RH, UK\\
$^{10}$ STFC Rutherford Appleton Laboratory, Chilton, Didcot, Oxfordshire OX11 0QX, UK\\
$^{11}$ Centre for Astrophysics Research, Science and Technology Research Institute, University of Hertfordshire, Hatfield AL10 9AB, UK\\
$^{12}$ McDonald Observatory, University of Texas at Austin, 1 University Station C1402, Austin, TX 78712, USA\\
$^{13}$ Institute for Astronomy, University of Edinburgh, Royal Observatory, Edinburgh EH9 3HJ, UK\\
}
\begin{document}

\date{}

\pagerange{\pageref{firstpage}--\pageref{lastpage}} \pubyear{2002}

\maketitle

\label{firstpage}

\begin{abstract}
We present a stellar mass-metallicity relation at $z\sim1.4$ with an unprecedentedly large sample of $\sim340$ star-forming galaxies obtained with FMOS on the Subaru Telescope. We observed K-band selected galaxies at $1.2\le z_{ph} \le 1.6$ in the SXDS/UDS fields with $M_{*}\ge 10^{9.5} M_{\sun}$, and expected F(H$\alpha$) $\ge$ $5\times 10^{-17}$ erg s$^{-1}$ cm$^{-2}$. Among the observed $\sim1200$ targets, 343 objects show significant H$\alpha$ emission lines. The gas-phase metallicity is obtained from [N\,{\sc ii}]$\lambda$6584/H$\alpha$ line ratio, after excluding possible active galactic nuclei (AGNs). Due to the faintness of the [N\,{\sc ii}]$\lambda$6584 lines, we apply the stacking analysis and derive the mass-metallicity relation at $z\sim1.4$. Our results are compared to past results at different redshifts in the literature. The mass-metallicity relation at $z\sim1.4$ is located between those at $z\sim0.8$ and $z\sim2.2$; it is found that the metallicity increases with decreasing redshift from $z\sim3$ to $z\sim0$ at fixed stellar mass. Thanks to the large size of the sample, we can study the dependence of the mass-metallicity relation on various galaxy physical properties. The average metallicity from the stacked spectra is close to the local FMR in the higher metallicity part but $\ga0.1$ dex higher in metallicity than the FMR in the lower metallicity part. We find that galaxies with larger $E(B-V)$, $B-R$, and $R-H$ colours tend to show higher metallicity by $\sim0.05$ dex at fixed stellar mass. We also find relatively clearer size dependence that objects with smaller half light radius tend to show higher metallicity by $\sim0.1$ dex at fixed stellar mass, especially in the low mass part.
\end{abstract}


\section[]{Introduction\label{sec:introduction}}
Heavy elements are synthesized in stars and returned into the interstellar medium (ISM), from which stars of new generation form, reflecting the result of the past star-formation activity in a galaxy. Thus, the gas-phase metallicity (hereafter metallicity) is a key parameter in understanding the processes of the formation and the evolution of a galaxy, though this process is somewhat complicated because of the effects of gas flow process such as gas inflow and outflow. It is known that the metallicity of galaxies correlates with their total mass \citep{Lequeux:1979p18393}. With the extent works in the local universe, the relation between stellar mass and metallicity (hereafter mass-metallicity relation) is well established; \citet{Tremonti:2004p4119} found the clear mass-metallicity relation at $z\sim0.1$ with a large sample of $\sim53\ 000$ Sloan Digital Sky Survey (SDSS) galaxies \citep{Abazajian:2004p27347}.

Cosmological evolution of the mass-metallicity relation is important to reveal the galaxy evolution. At $z\sim0.7-0.8$, \citet{Savaglio:2005p3325} with a sample of $\sim60$ galaxies, and recently \citet{Zahid:2011p11939} with a larger sample of $\sim1300$ galaxies show the downward shift of the mass-metallicity compared to the local one by \citet{Tremonti:2004p4119}. This general trend has been seen at stellar masses ranging from $10^{8}$ M$_{\odot}$ to $10^{11}$ M$_{\odot}$ \citep[e.g.,][]{Henry:2013p25750}. By using $\sim3000$ galaxies at $z\sim0.1-0.8$, \citet{Moustakas:2011p25716} examine the evolution of the mass-metallicity relation systematically, and find that the relation shifts towards lower metallicity with increasing redshift without changing its shape from $z\sim0.8$ to $z\sim0$. At higher redshift, by using a sample of $\sim90$ galaxies at $z\sim2.2$, \citet{Erb:2006p4143} found that the mass-metallicity relation shifts downward by $0.56$ dex from that at $z\sim0.1$ by \citet{Tremonti:2004p4119}. At $z\sim3$, \citet{Maiolino:2008p5212} and \citet{Mannucci:2009p8028} found the further downward shifts by using $\sim20$ galaxies, suggesting the smooth evolution of the mass-metallicity relation from $z\sim3$ to $z\sim0$ (see also \citealt{Zahid:2013p25847} and references therein).

The mass-metallicity relation at $z=1-2$, however, still remains unclear. The mass-metallicity relations with higher metallicity at fixed stellar mass than that by \citet{Erb:2006p4143} are reported \citep{Hayashi:2009p4235,Yoshikawa:2010p4286,Onodera:2010p4273}. One of the reason for this discrepancy may be the smallness of the sample size for these works; the sample size in each work is $\sim10-20$, which are $5-10$ times smaller than that by \citet{Erb:2006p4143}. Another reason for the discrepancy may be the differing selection methods among these samples. The redshift range of $z\sim1-2$ is an important phase in the evolutionary history of galaxies. Galaxies are in the most active phase in this redshift range; the cosmic star-formation density peaks at $z\sim2$ \citep[][]{Hopkins:2006p4539}. Thus, increasing the sample size and establishing the mass-metallicity relation at this redshift range is very crucial to understand the galaxy formation and evolution. 

It is also important to explore the physical drivers of the scatter in the mass-metallicity relation and their evolution with redshift. \citet{Tremonti:2004p4119} show that the mass-metallicity relation at $z\sim0.1$ has a scatter of $\sim0.1$ dex with roughly half of the scatter being attributable to observational error. The origin of the scatter, in other words, the dependence of the mass-metallicity relation on the other physical parameters must be an important clue to the understanding the galaxy formation and evolution. The parameter dependence on the mass-metallicity relation has been argued by various works: For instance, \citet{Tremonti:2004p4119} report that galaxies with higher stellar mass surface density tend to show higher metallicity at fixed stellar mass, suggesting the efficient transform from gas to stars raising the metallicity. \citet{Ellison:2008p7997} use the SDSS sample to show that at fixed stellar mass galaxies with larger sizes or higher specific star-formation rate (sSFR) have lower metallicities. Morphology dependence of the mass-metallicity relation is also reported: \citet{Rupke:2008p8273} find that (ultra) luminous infrared galaxies, which generally present interaction or merging features, tend to show systematically lower metallicity on the mass-metallicity relation. \citet{SolAlonso:2010p18793} also find that galaxies with the strongly disturbed morphology tend to show lower metallicity at fixed stellar mass. These results suggest the existence of the interaction- and/or merger-induced gas inflow from the external low metal region of galaxies.

Recently, the dependence of the mass-metallicity relation on star-formation rate (SFR) has been discussed \citep{Mannucci:2010p8026,LaraLopez:2010p16580,Yates:2011p16030} by using SDSS galaxies at $z\sim0.1$. \citet{Mannucci:2010p8026} find that at fixed stellar mass galaxies with higher SFR tend to show lower metallicity. With the SFR as the third parameters, they proposed the fundamental relation between the stellar mass, metallicity, and SFR (Fundamental Metallicity Relation; FMR); galaxies make a surface in the 3D-space at $z\la2.5$, suggesting that the evolution of the mass-metallicity relation is due to the apparent shift on this surface with changing SFRs. By using the zCOSMOS sample at $0.2<z<0.8$, \citet{Cresci:2011p20053} find the similar SFR dependence on the mass-metallicity relation following the FMR at $z\sim0.1$. At higher redshifts of $z\ga1$, however, dependence on the other parameters including the SFR on the mass-metallicity relation still remains unclear due to the smallness of the previous sample, requiring a larger sample to search for wider parameter space at high redshift.

With these motives, recently, we conduct near-infrared (NIR) spectroscopic surveys at $z=1-2$ by using the Fibre Multi-Object Spectrograph (FMOS), which is a fibre-fed type multi-object NIR spectrograph on the Subaru Telescope \citep{Kimura:2010p11396}. Up to 400 objects are observed simultaneously with 1.2 arcsec diameter fibres manipulated by a fibre positioner called ``Echidna'' on the prime focus (30 arcmin diameter FoV) of the Subaru Telescope. FMOS has two NIR spectrographs (IRS1 and IRS2) covering the wavelength range of $0.9-1.8$ $\mu$m, where the spectral resolution is typically $R\sim650$ in the \textit{low resolution} (LR) mode and $R\sim3000$ in the \textit{high resolution} (HR) mode. FMOS equips the OH-suppression system; the strong OH airglow emission lines in HR spectra are removed by the mask mirrors. The HR spectra, in which the OH-lines are removed, is degraded by a VPH grating in the LR mode. The detailed descriptions of the OH-suppression system can be found in Section 2.4 of \citet{Kimura:2010p11396}.

The initial results from the survey are already published by \citet{Yabe:2012p23184} (hereafter Y12). They found the mass-metallicity relation by using a sample of $\sim70$ star-forming galaxies (SFGs) at the redshift range of $1.2\le z\le 1.6$ with a median of $z\sim1.4$, and found that the mass-metallicity relation evolves smoothly from $z\sim3$ to $z\sim0$ by compiling other works at various redshifts. They also found that there exists intrinsic scatter of $\ga0.1$ dex in the mass-metallicity relation at $z\sim1.4$. The relatively larger sample than previous studies at the similar redshift allows us to examine the dependence of the mass-metallicity relation on various physical parameters: We found trends that galaxies with the higher SFR and larger half light radius ($r_{50}$) show lower metallicity at fixed stellar mass. The sample size in the initial work, however, is still limited to reveal the parameter dependence on the mass-metallicity relation. In this paper, we present the results from the subsequent surveys with the Subaru/FMOS for establishing the mass-metallicity relation at $z\sim1.4$ with $\sim5$ times larger sample than our previous work, which is the largest galaxy sample ever at $z>1$. 

Throughout this paper, we adopt the concordance cosmology with $(\Omega_{M},\Omega_{\lambda},h)=(0.3,0.7,0.7)$. All magnitudes are in the AB system \citep{Oke:1983p15127}.

\begin{figure*}
\begin{center}
\includegraphics[scale=0.50,angle=270]{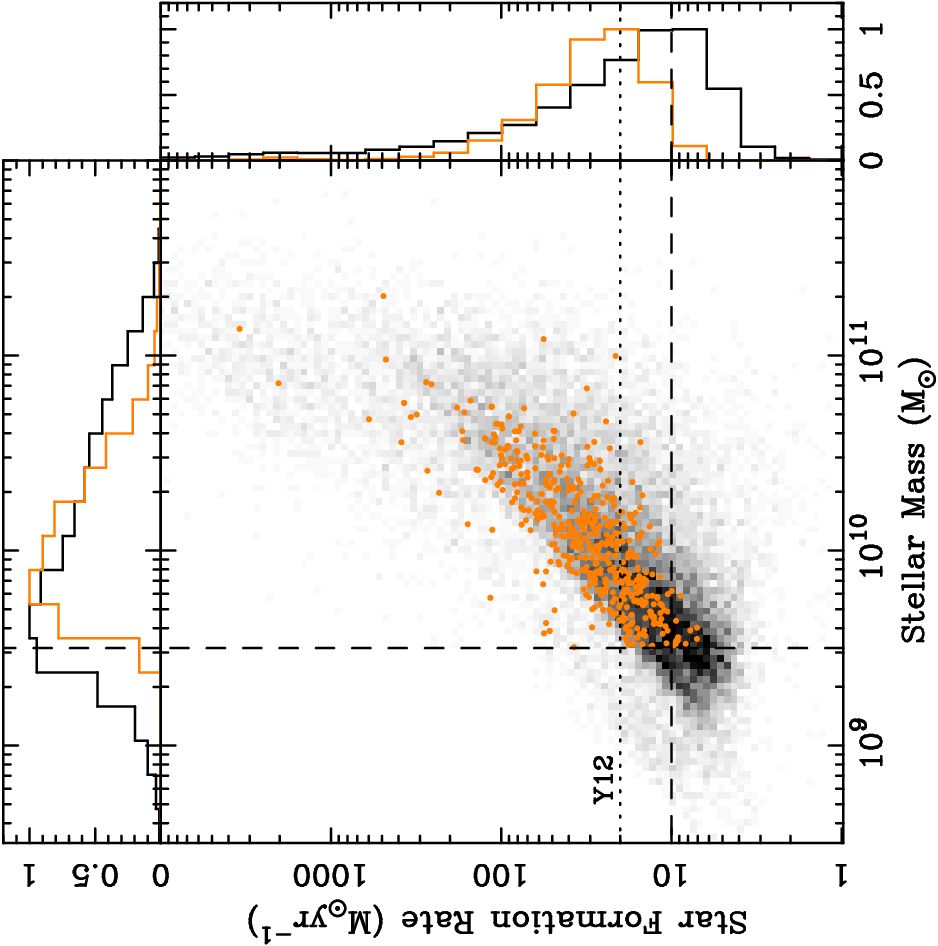}
\includegraphics[scale=0.50,angle=270]{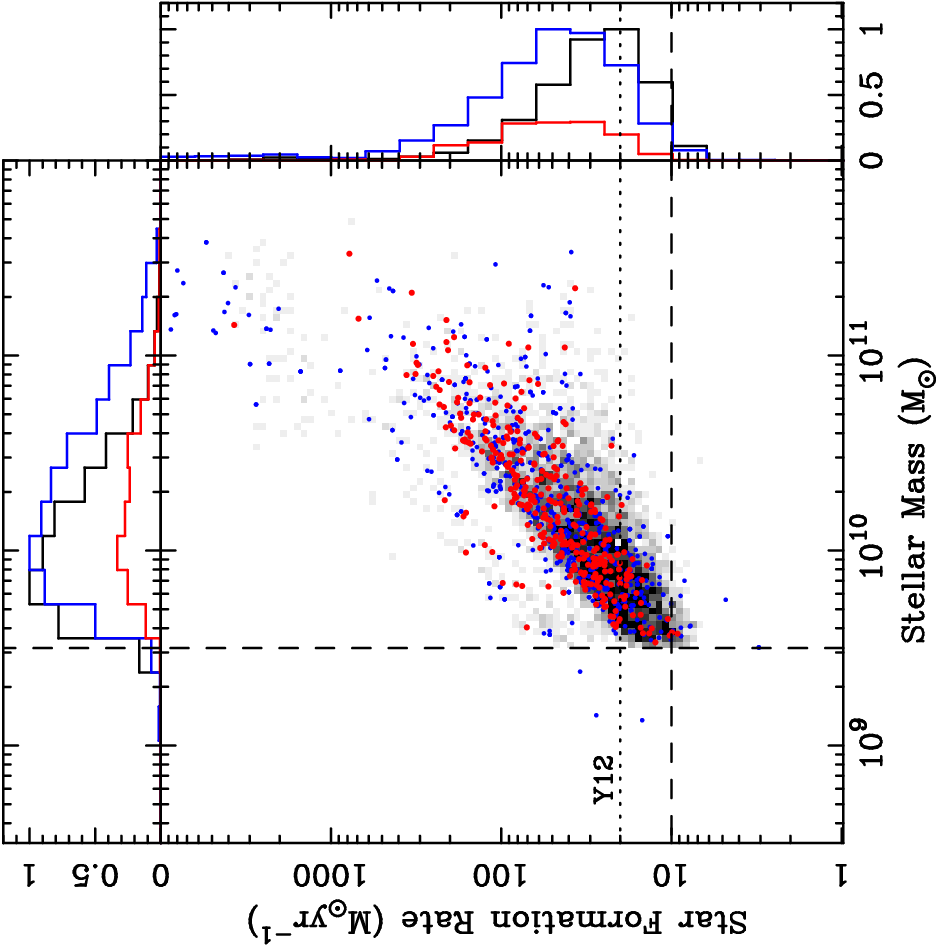}
\caption{The galaxy samples on the $M_{*}$$-$SFR diagram. \textbf{Left}: The primary sample of 19144 objects with $K_{s} < 23.9$ mag, $1.2 \le z_{ph} \leq 1.6$ is indicated as gray scale map. For clarity, 500 objects randomly selected form our target sample with $M_{*} \ge 10^{9.5}M_{\sun}$ and the expected F(H$\alpha$)$>$5$\times$10$^{-17}$ erg s$^{-1}$ cm$^{-2}$ is indicated by orange dots. The peak-normalized distributions of the primary sample (black line) and the target sample (orange line) against the each axis are presented in sub-panels. \textbf{Right}: The target sample of 4745 objects is indicated as gray scale map. The actually observed objects in the FMOS observations are indicated by blue dots. Some objects with stellar mass less than $10^{9.5}M_{\sun}$ were observed as test cases. Objects with H$\alpha$ detection with SN of $\ge$3 are indicated by red dots. Details about the H$\alpha$ detections are described in Section \ref{sec:reductions}. In the sub-panels, the peak-normalized distribution for the targets (black line) and the observed sample (blue line) are presented. The peak of the distribution of the H$\alpha$ detection sample (red line) is scaled by using the ratio of the H$\alpha$ detection sample to the observed sample. In both panels, the vertical dashed line shows the stellar mass limit of $10^{9.5}M_{\sun}$ and the horizontal dashed line indicates SFR of 10 M$_{\sun}$ yr$^{-1}$. The SFR limit of the sample by Y12 is also presented as dotted line for reference.\label{fig:Sample}}
\end{center}
\end{figure*}

\section[]{Sample and Observations\label{sec:sample}}
\subsection{K-band Selected Galaxy Sample in the SXDS/UDS field\label{sec:PrimarySample}}
The galaxy sample in this work is selected from the K-band detected galaxy catalogue in the overlapped region of the Subaru XMM-Newton Deep Survey (SXDS) and the UKIRT Infrared Deep Sky Survey/Ultra Deep Survey (UDS) (hereafter SXDS/UDS) with broad band Spectral Energy Distributions (SEDs) from far-UV to mid-IR (MIR). For the optical data, we use the public data release DR1 \citep{Furusawa:2008p15159} of the SXDS Subaru/Suprime-Cam images ($B$, $V$, $R_{C}$, $i'$, and $z'$-band). For NIR data, we use the UKIDSS release DR8 \citep{Lawrence:2007p15348} of the UDS UKIRT/WFCAM images ($J$, $H$, and $K_{s}$). For the mid-IR data, Spitzer/IRAC images (3.6 $\mu$m, 4.5 $\mu$m, 5.8 $\mu$m, and 8.0 $\mu$m) are taken from the SpUDS Spitzer Legacy Survey (Dunlop et al. in prep.). The detailed information such as limiting magnitudes of the data are described by Y12. In addition to these data, we use the far-UV and near-UV data taken from \textit{GALEX} archived image (GR6) and $U$-band images taken from the CFHTLS wide surveys. These supplemental data, however, are relatively shallow compared to the optical to MIR data and do not affect resulting photometric redshifts (phot-$z$s) and other physical parameters derived from the SED fitting. The optical to NIR images are aligned to the $K_{s}$-band images and convolved so that their PSF FWHMs are matched to $0.91$ arcsec which is the worst PSF size in $H$-band. The IRAC images are also aligned to the $K_{s}$-band images but their PSFs were not matched to the optical to NIR images.

Object detection and photometry are done by using \textsc{SExtractor} \citep{Bertin:1996p15563} with \textit{double image} mode for the aligned and PSF-matched optical to NIR images. We detected $\sim$190\,000 objects down to $K_{s}\sim25.5$ mag in the whole SXDS/UDS area of $\sim$ 2400 arcmin$^{2}$. In each image and object, the total magnitude is derived from the aperture magnitude with 2.0 arcsec aperture that is scaled to the \texttt{MAG\_AUTO} in the $K_{s}$-band image. For IRAC MIR images, the detection and the photometry is also carried out by the \textit{double image} mode of the SExtractor. The total magnitude are calculated from the aperture magnitude with 2.4 arcsec aperture by applying aperture corrections. The detailed descriptions of the detection and photometry are presented by Y12. 

Phot-$z$s of the detected objects with the broad band SED from \textit{GALEX} FUV to \textit{Spitzer} MIR are determined by using \textsc{Hyperz} \citep{Bolzonella:2000p243}. The phot-$z$s well agree with the resulting spectroscopic redshift (spec-$z$) in the literature \citep[Akiyama et al. 2013, in prep.;][]{Simpson:2012p25768, Smail:2008p15567} and of 343 H$\alpha$ detected objects in our observations, which is described in Section \ref{sec:SpectralFitting}. The phot-$z$s, however, tend to be systematically smaller than the spec-$z$ by $\Delta z \sim0.05$ at $z=1-2$. The standard deviation of the difference is $\sigma \sim0.05$ after removing the systematics. Although the offset is presumably due to the phot-$z$ code, further investigation on the origin is not carried out in this work.

The stellar masses are derived by performing SED fitting by using the \textsc{SEDfit} \citep{Sawicki:2012}. We found that the effect of the systematic uncertainty of the phot-$z$ determination on the stellar mass and also the sample selection is negligible: The stellar mass recalculated by using the phot-$z$ corrected for the systematics is larger than the original stellar mass by only $5.7\pm1.9\%$. In the sample selection below, we use the phot$-z$ which is not corrected for the systematics. As we mention in Section \ref{sec:MZR}, however, we use the physical properties such as the stellar mass that are recalculated by using the spec-$z$ after Section \ref{sec:MZR}.

The colour excesses are estimated from the rest-frame UV colours in a manner similar to that of \citet{Daddi:2004p6750} and \citet{Daddi:2007p1460}. The SFR is derived from the rest-frame UV luminosity density by using the conversion by \citet{Kennicutt:1998p7465}. The rest-frame UV luminosity density is calculated from the observed magnitude in $B$-band, covering the rest-frame wavelength of $1500-2300 $\AA\ at $z=1-2$, which is within the valid wavelength range of the conversion. The intrinsic SFR is derived by correcting for the extinction with the $E(B-V)$ derived above assuming the Calzetti extinction curve \citep{Calzetti:2000p7012}.

In order to measure the metallicity, we aim to target galaxies with detectable nebular emission. We therefore estimate the H$\alpha$ flux of each galaxy (hereafter ``the expected H$\alpha$ flux'') from the intrinsic SFR and the $E(B-V)$ described above. Since it is suggested that the extinction is significantly larger for the ionized gas than for the stellar component \citep{Calzetti:2000p7012}, we convert the obtained $E(B-V)$ to that for the ionized gas by using a prescription by \citet{CidFernandes:2005p3252}. For the conversion from the SFR to the H$\alpha$ luminosity, we use the relation by \citet{Kennicutt:1998p7465}. Although there exists a large scatter, the expected H$\alpha$ flux roughly agrees with the actually observed H$\alpha$ flux in the FMOS observations in the range from $5\times10^{-17}$ to $1\times10^{-15}$ erg s$^{-1}$ cm$^{-2}$. Since the expected H$\alpha$ flux is derived by assuming the extra reddening for the emission line than for the stellar component by using the prescription by \citet{CidFernandes:2005p3252}, this result indirectly supports the possibility of the differing extinction between the emission line region and the stellar components. Details on comparison of the expected and observed H$\alpha$ flux is presented in Section \ref{sec:SpectralFitting}.

In order to check the possibility of the AGN contamination, the sample is cross-correlated with X-ray sources in the SXDS \citep{Ueda:2008p13177} and objects cross-matched within the error circle of the X-ray source are excluded from the sample. Thus, the X-ray bright AGNs (L$_{X(2-10\textrm{keV})}$$\ga$10$^{43}$ erg s$^{-1}$) are excluded from our sample. By this process, $\sim 1 \%$ of the target sample, which is described in the next subsection, is excluded as possible AGNs.

\subsection{Target Sample\label{sec:TargetSample}}
For the spectroscopic observations with FMOS, the sample is constructed from the K-selected catalogue described above with the following selection: $K_{s} < 23.9$ mag, $1.2\le z_{ph}\le 1.6$, and $M_{*}\ge 10^{9.5} M_{\sun}$, which are the same criteria as Y12. In Figure \ref{fig:Sample}, intrinsic SFRs derived from the rest-frame UV are plotted against the stellar mass (hereafter $M_{*}-SFR$ diagram) for the samples; the distribution along each axis is also presented in sub-panels. In the left panel, all galaxies with $K_{s} < 23.9$ mag, $1.2\le z_{ph}\le 1.6$ are plotted as a gray-scale map. It is shown that the SFR correlates well with the stellar mass, which has been reported by various studies \citep[e.g.,][]{Daddi:2007p1460}. The number of galaxies in this primary sample is 19144, which is too large to observe in the limited observing time; also, many of the primary sample may show very faint H$\alpha$ emission. Thus in order to make an efficient survey, we construct an (expected) flux limited sample as the target sample.

The target sample is selected from the primary sample by the expected H$\alpha$ line flux; the method of the calculation is described in the previous subsection. Although we mainly target bright sample with F(H$\alpha$)$^{exp}$$\ge$1.0$\times$10$^{-16}$ erg s$^{-1}$ cm$^{-2}$ in the initial observing runs (Y12), we also target fainter sample with F(H$\alpha$)$^{exp}$$\ge$5.0$\times$10$^{-17}$ erg s$^{-1}$ cm$^{-2}$ in the subsequent FMOS observations. The number of galaxies in the target sample with F(H$\alpha$)$^{exp}$$\ge$1.0$\times$10$^{-16}$ erg s$^{-1}$ cm$^{-2}$ and F(H$\alpha$)$^{exp}$$\ge$5.0$\times$10$^{-17}$ erg s$^{-1}$ cm$^{-2}$ is 1574 and 4745, respectively. In the left panel of Figure \ref{fig:Sample}, our target sample is plotted by orange dots (for clarity, 500 galaxies randomly selected from the sample are plotted). The distribution of the target sample appears to be similar to that of the original sample (gray-scale map). It is also shown that our target sample mostly covers SFR of $\ga 10 M_{\sun}$yr$^{-1}$, which is $\sim2$ times smaller than the limit by Y12.

The fibre configuration design for the target objects in each FMOS field of view (FoV) was done by using the FMOS fibre allocation software\footnote{Details of the allocation software can be found at http://www.naoj.org/Observing/Instruments/FMOS/observer.html}, in which the fibre configuration is optimized semi-automatically by referring the allocation priority. Although we target sample with F(H$\alpha$)$^{exp}$$\ge$5.0$\times$10$^{-17}$ erg s$^{-1}$ cm$^{-2}$, we gave higher priorities in the fibre configuration to objects with F(H$\alpha$)$^{exp}$$\ge$1.0$\times$10$^{-16}$ erg s$^{-1}$ cm$^{-2}$. The number of the allocated objects with F(H$\alpha$)$^{exp}$$\ge$1.0$\times$10$^{-16}$ erg s$^{-1}$ cm$^{-2}$ and F(H$\alpha$)$^{exp}$$\ge$5.0$\times$10$^{-17}$ erg s$^{-1}$ cm$^{-2}$ is 973 and 1209, respectively. In the right panel of Figure \ref{fig:Sample}, our target sample is indicated as a gray-scale map and the actually observed sample is plotted by dots. We indicate H$\alpha$ detections by red dots. Details of the H$\alpha$ detection are described in Section \ref{sec:SpectralFitting}. Some objects with the stellar mass of $\le 10^{9.5} M_{\sun}$ were observed in the FMOS observations as test cases. It appears to be shown that the distribution of our observed sample is similar to that of the target sample. However, the distribution of the observed sample is somewhat biased towards larger stellar mass and SFR. We placed the higher priority on a galaxy with the larger expected Ha$\alpha$ flux; such a sample consists of objects with relatively large stellar mass and SFR. As we mention in Section \ref{sec:MZR_dependence}, there is no clear correlation between SFR and metallicity at fixed stellar mass. Therefore, the effects of the selection bias in the observations on the final results are considered to be small.

\subsection{Observations\label{sec:observations}}
The observations were carried out in the FMOS guaranteed time observing runs, engineering observing runs for science verification, and open use observations. The observing runs in 2010 are described by Y12. The observations in 2011 runs were carried out on 2011 October $7-15$, December $1-2$, and $15-18$ under various weather conditions. The typical seeing size measured with the Echidna sky-camera during the observations was 0.9 arcsec in $R$-band. 

The observations were all made with the Cross Beam Switch (CBS) mode. In the CBS mode, two fibres were allocated for one target and the sky. In an exposure, one fibre looks at the target and the other looks at the sky typically $60-90$ arcsec away from the target (this configuration is referred as Pos. A). In the next exposure, by nodding the telescope with the separation of the two fibres, the fibre for the target in the previous exposure looks at the sky and that for the sky looks at the target in turn (Pos.B). After the set of the exposures, the telescope moves back to the Pos. A. In the typical observing sequence for one field of view, after the first configuration of the fibre spine ($15-20$ min.) at the beginning, an exposure for Pos. A was made (15 min.), and we moved to the Pos. B for the next exposure (15 min.). After a set of exposures (Pos. A and B) was made, we tweak the fibre spine configuration ($10-15$ min.) and then a set of Pos. A and B was made again. We ran over the sequence until the required integration time was achieved. The focussing check was made every 1-2 hours. The typical exposure time was $\sim3-4$ hours on source with the observing time of $\sim5-6$ hours for one field of view. The typical positional error of the allocated fibres during the observations is $\sim0.2$ arcsec.

In order to cover wide wavelength coverage, we employed a spectral set up of LR mode in both spectrographs of IRS1 and IRS2. The LR mode covers $0.9-1.8$ $\mu$m simultaneously with the typical spectral resolution of $R\sim650$ at $\lambda\sim1.30$ $\mu$m, which are measured from the Th-Ar lamp frames. The pixel scale in the LR-mode is 5 \textrm{\AA}.

\section{Data Reduction and Analysis}
\subsection{Data Reduction\label{sec:reductions}}
The obtained data were reduced with the FMOS pipeline \textsc{FIBRE-pac} and detailed descriptions for the pipeline are presented by \citet{Iwamuro:2011p18754}. The process of the reduction is almost the same as that by Y12. Here we briefly describe the outline of the process. For a set of exposures, the $A-B$ sky subtraction is carried out. For the obtained 2D spectra, the distortion correction is done by tracing the 2D spectra of the dome flat images. Although major OH emission lines are removed by the OH suppression system, a fraction of the OH lines remains due to several reasons such as the deviation of the OH masks and the time variation of the OH lines themselves. The residual sky subtraction is also carried out in the reduction pipeline \citep[see Figure 5 and 6 of][]{Iwamuro:2011p18754}. In the typical observations, we obtained $6-8$ pairs of the $A-B$ frames and the obtained spectra are combined for the total exposures. Since the spectrum of one target is obtained by two fibres in the CBS mode, the final spectrum of one target is obtained by merging the CBS pair spectra. The wavelength calibration is done by using Th-Ar lamp frames. The uncertainty of the wavelength calibration is $\Delta \lambda \sim 5$ \textrm{\AA}.

The relative flux calibration was done by using several F, G or K-type stars selected based on $J-H$ and $H-K_{s}$ colours and observed simultaneously with other scientific targets. The uncertainty of the relative flux calibration in the determination of the star type from the observed spectral slope is estimated to be $\sim10\%$ \citep[see Fig. 11 of][]{Iwamuro:2011p18754}. The absolute flux is determined from the observed count rate by assuming the total throughput of the instrument, which is calibrated by using the moderately bright stars ($\sim$16 mag in $J$- or $H$-band) in the best weather conditions (clear sky with seeing of $\sim$ 0.7 arcsec) in previous FMOS observations. For the absolute flux, we correct the degradation of the atmospheric transmission due to the poor weather by using the ratio of the observed spectral flux and the flux from the broad band photometry \citep[see Figure 14 of][]{Iwamuro:2011p18754}. Although the uncertainty of the absolute flux calibration, itself, is $\sim10\%$, there may be uncertainties of $20-30\%$ that come from a variation of the atmospheric transmission and the seeing conditions. It is worth noting that the uncertainty of the absolute flux does not affect the relative physical quantity such as line ratios close in wavelength and thus metallicity. Finally, 1D spectra are extracted from the 2D spectra after the wavelength and flux calibration.

\begin{figure}
\includegraphics[angle=-90, scale=0.50]{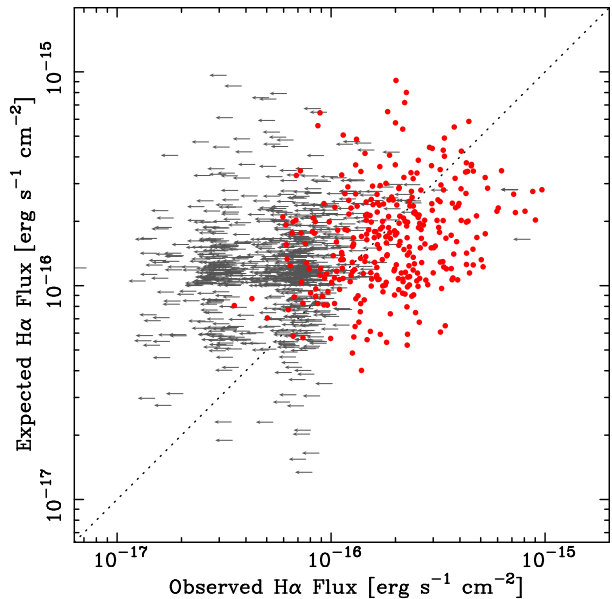}
\caption{Comparison of the observed H$\alpha$ flux and the expected flux. H$\alpha$ detected objects are indicated by red filled circles. Objects with H$\alpha$ non-detection are presented as upper limits for the observed flux. The equivalent line is indicated by dotted line. \label{fig:HaFluxComparison}}
\end{figure}

\subsection{Spectral Fittings and Line Flux Measurements\label{sec:SpectralFitting}}
In the observations, $\sim$1200 objects were observed in total and 343 objects show a significant emission line feature in H-band ($\lambda\sim1.4-1.7$ $\mu$m) according to our eye inspections; automatic detection of the emission line for a large survey data by using FMOS is being now investigated (Tonegawa et al. 2013, in prep.). Multiple emission lines such as [N\,{\sc ii}]$\lambda$6584 in H-band and [O\,{\sc iii}]$\lambda$5007 and H$\beta$ in J-band can be also seen in some cases. If there is one emission line feature in the H-band, there is a possibility that the emission line is an [O\,{\sc ii}]$\lambda3727$ at $z=2.76-3.56$. According to the comparison of the phot$-z$ and spec$-z$ in Section \ref{sec:PrimarySample}, however, only $\sim0.4\%$ of objects with the phot-$z$ of $z=1.20-1.60$ have the spec-$z$ of $z=2.76-3.56$. Thus, we conclude that all these objects are likely to be galaxies at $z=1.2-1.6$.

For these 343 objects, the fluxes of [N\,{\sc ii}]$\lambda$6584, H$\alpha$, [O\,{\sc iii}]$\lambda\lambda$4959,5007, H$\beta$ emission lines are measured by fitting the reduced one-dimensional spectra. Since the OH-masks in FMOS remove strong OH airglow emission lines as well as a part of the emission line from the target object, we employ a complicated fitting process. In this work, we use the same method as that by Y12, which includes the effect of the OH-masks in the fitting process. The detailed description of the fitting method and its limitation are presented in section 3.1 of Y12. To summarize briefly, we use a model emission line spectrum with the Gaussian profile, taking redshift, line width, and normalization as free parameters. The flux density of the spectrum at the wavelength covered by the OH-mask is reduced to be zero. Then the spectrum is degraded to the spectral resolution of $R\sim650$ in the LR mode. Since the observed spectrum is corrected for the effects of the OH-mask on the continuum light in the data reduction process, this effects is also taken into consideration for the model spectrum. Then, the corrected LR model spectrum is fit to the observed data to constrain the free parameters of redshift, line width, and the flux normalization. The schematic view of spectra at each stage can be found in Figure 3 of Y12. In this method, the uncertainty of the flux recovery depends on how much flux the OH masks remove (see Figure 4 of Y12).

In the fitting process of the method with including the OH-mask effects, flux loss rate $\mu^{loss}$ is quantified as $\mu^{loss}\equiv f^{loss} / f^{int}$, where $f^{loss}$ and $f^{int}$ are the flux lost by the OH-mask and the intrinsic flux calculated from the best-fitting spectral model, respectively. As we mentioned above, with increasing $\mu^{loss}$, the uncertainty of the recovering flux in the fitting process increases. In this paper, we use only objects with a threshold of $\mu^{loss} \le 67\%$ for all emission lines, which is trade-off between the accuracy of the recovering flux and the sample size. This criterion decreases the sample size to 305 out of 343 objects, but also decreases the uncertainty as to the flux recovery down to $\la10\%$. Various results in this paper do not change largely if the threshold is taken to be 50\%.

The fluxes and other properties are also derived by using the simple fitting method without including the OH-mask effects. The observed line fluxes by using these two methods agree with each other within $\sigma\sim10\%$ without a systematic difference at the flux level of $\ga 1\times10^{-17}$ erg s$^{-1}$ cm$^{-2}$. The observed spec-$z$ also agrees within $\Delta z\sim0.0002$ by the two methods. The details on the effects of the OH-mask on the observed spectra and the two fitting methods are described in Section 2.3 of Y12. In this work, we use the former method that includes the OH-mask effects as the fiducial fitting method unless otherwise noted.

The signal-to-noise ratio (S/N) is estimated from the observed line flux, which is not corrected for the mask loss, and the noise level measured from the fluctuation of continuum in a wavelength window of $\pm$0.1 $\mu$m from the emission line. The estimation of the noise level is more practical than that obtained through the FIBRE-pac pipeline, where the noise level is idealized. For the line detection, here we use the threshold of line S/N higher than 3.0, unless otherwise noted. For 343 objects among $\sim1200$ targets, the H$\alpha$ emission line was detected with the S/N larger than 3.0.

Since the aperture size of the FMOS fibre of 1.2 arcsec diameter is generally smaller than the typical observed size of target galaxies at the redshifts even if in a good seeing condition, a part of the light from the target object is lost. Although this aperture effect on the relative quantities such as a line ratio and therefore the metallicity is relatively small, it is critical to absolute quantities such as the total H$\alpha$ flux and therefore the SFR. We recover the flux loss due to the aperture effect by using the same method presented by Y12. In this method, the amount of the flux loss by the fibre aperture is estimated from the $r_{50}$ of the target, which is determined from the WFCAM $K-$band image by deconvolving the typical PSF size, and the seeing size in the observations. Here, we assume that the position of the fibre centre during the observations coincides with the centre of the target galaxy. The observed line fluxes are all corrected by using the amount of the flux loss; the typical correction factor is a factor of $\sim 2$. The detailed procedure of recovering the aperture loss and the possible uncertainty are presented in Section 3.3 of Y12.

\begin{figure*}
\includegraphics[angle=-90, scale=1.00]{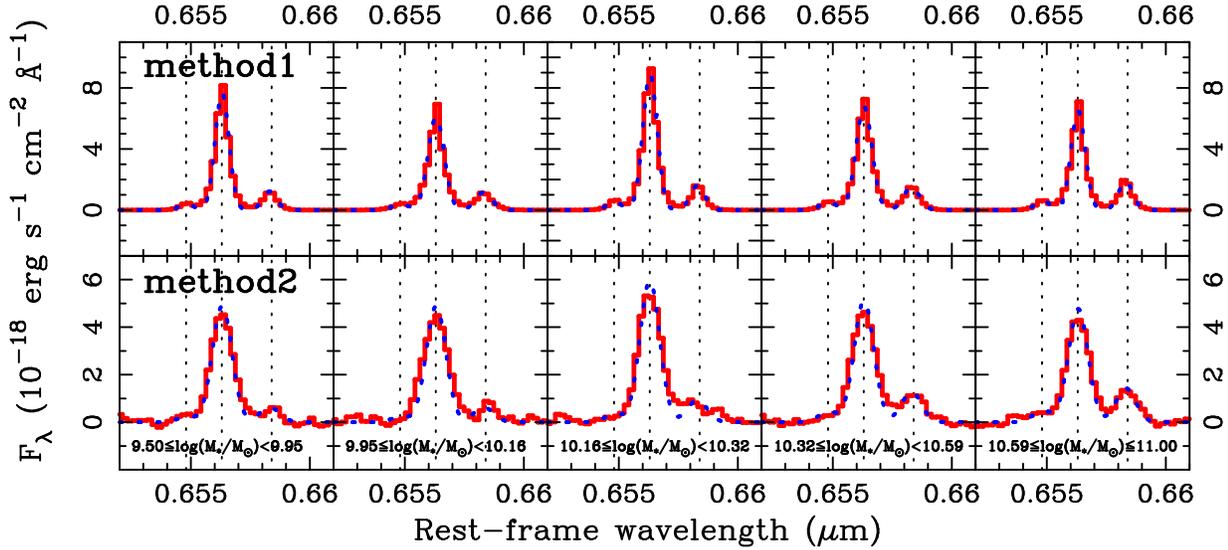}
\caption{Stacked spectra (red solid histogram) and the best-fitting models (blue dotted line) in five stellar mass bins. The spectra obtained from method 1 (stacking the best-fit model spectra) and method 2 (stacking the observed spectra) are presented in the top and bottom sub-panels, respectively. The stellar mass ranges are presented in the figure. Vertical dotted lines indicate the positions of [N\,{\sc ii}]$\lambda$6548, H$\alpha$, [N\,{\sc ii}]$\lambda$6584 from left to right, respectively. \label{fig:Spectra}}
\end{figure*}

\begin{figure*}
\includegraphics[angle=-90, scale=0.50]{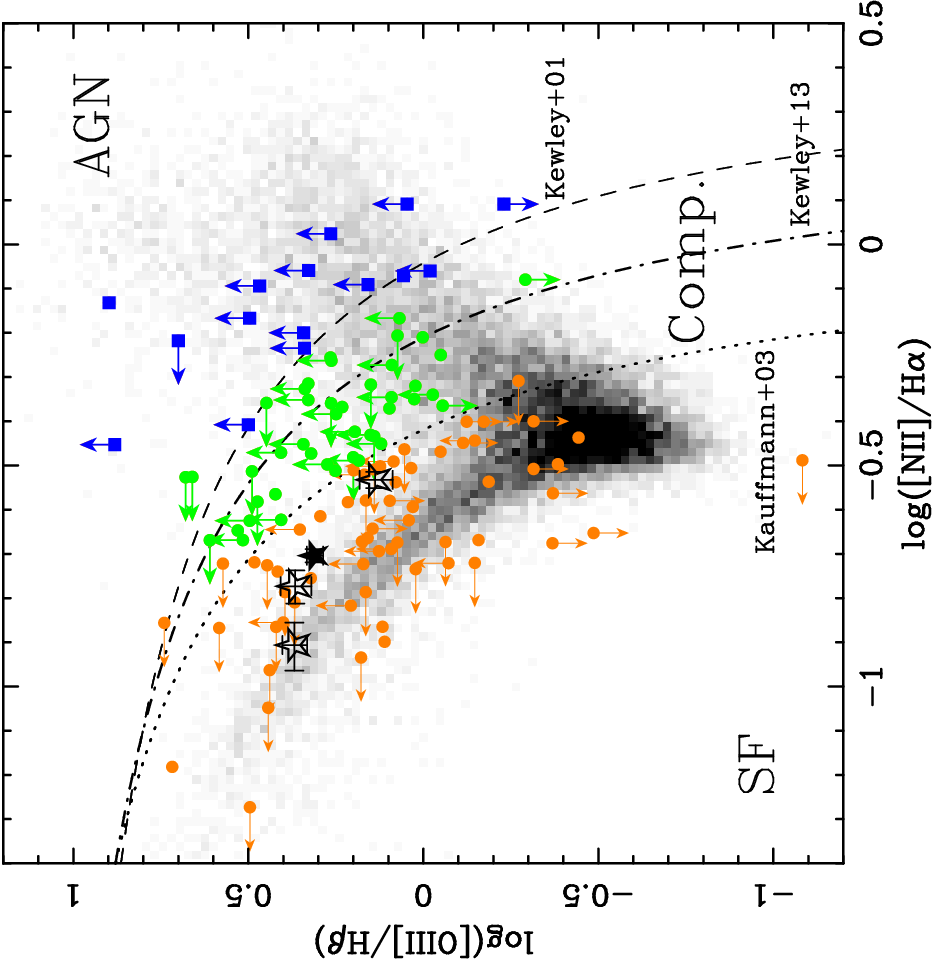}
\includegraphics[angle=-90, scale=0.50]{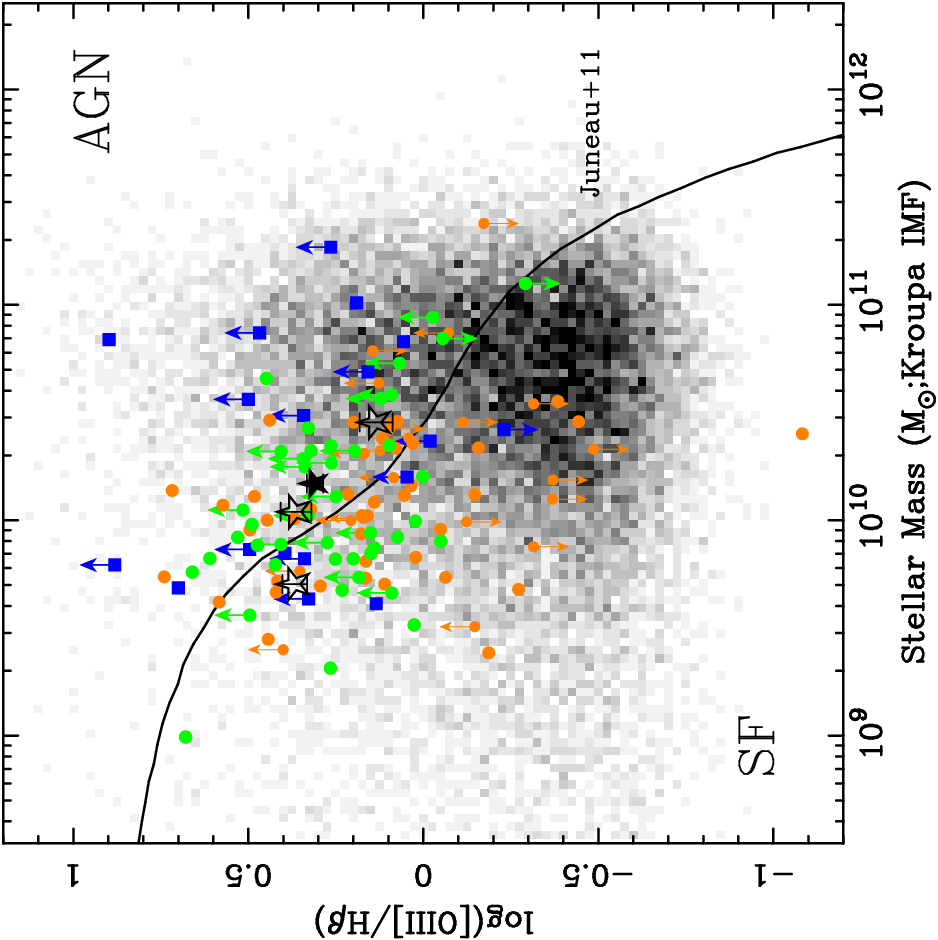}
\caption{\textbf{Left}: BPT diagram. Upper and lower limits are $2.5\sigma$ values. The result of the stacking analysis is shown by a black filled star for all sample but excluding the AGN candidates and open stars for three stellar mass bins. For comparison, local SDSS galaxies are plotted as gray scale map. The empirical criterion to separate the AGN and SF by \citet{Kauffmann:2003p8323} and the maximum theoretical line of starbursts by \citet{Kewley:2001p5158} are shown as dotted and dashed lines, respectively. The recent theoretical prediction at $z\sim1.4$ by \citet{Kewley:2013p25967} is presented by dot-dashed line. Objects identified as AGN, composite, and SF galaxies are plotted by blue, green, and orange symbols, respectively. \textbf{Right}: Stellar mass vs. [O\,{\sc iii}]$\lambda$5007 (MEx) diagram. Observed data points are presented by filled circles and arrows. The results of the stacking analysis are also presented as filled and open stars. For comparison, local SDSS galaxies are also plotted as gray scale map. The solid line shows a criterion for the AGN-SF separation by \citet{Juneau:2011p17996}. Objects identified in the BPT diagram (left panel) as AGN, composite, and SF galaxies are plotted by blue, green, and orange symbols, respectively. All the stellar mass is converted to the Kroupa IMF. \label{fig:BPT}}
\end{figure*}

The Balmer absorption could make a significant contribution to our estimation of H$\alpha$ and H$\beta$ flux \citep[e.g.,][]{Groves:2012p25816}. The effect of the stellar absorption on the H$\alpha$ and H$\beta$ emission lines is examined by using the stellar population synthesis models by \citet{Bruzual:2003p6157} with various stellar age and the star-formation history. The stellar ages of 100 Myr $-$ 1 Gyr tend to show the large absorption at any star-formation history. The largest rest-frame equivalent widths of H$\alpha$ and H$\beta$ are $\sim-5$ \AA\ and $\sim-10$ \AA\, respectively. The contribution to the median value of the obtained H$\alpha$ and H$\beta$ emission line fluxes are $\la5\%$ and $\la20\%$, respectively. Although the effect of the H$\alpha$ on results is very low, which is also reported by Y12, the H$\beta$ could affect the AGN rejection, which is described in Section \ref{sec:AGN} in details. However, the sample with rejecting the possible AGN and the subsequent results does not change, because only a few objects that were selected as AGN candidates becomes SF candidates even if we take account of the maximum H$\alpha$ and H$\beta$ absorption. Correction of the absorption for individual object is hard, because the uncertainty of the determination of the stellar age and the star-formation history from the SED fittings is very large. We thus do not apply the absorption correction for the obtained H$\alpha$ and H$\beta$ flux.

The detection rate of H$\alpha$ is $\sim30\%$; no strong dependence of stellar mass and intrinsic SFR on the detection rate can be seen, as shown in the right panel of Figure \ref{fig:Sample}. Comparison between the observed H$\alpha$ flux and the expected flux is presented in Figure \ref{fig:HaFluxComparison}. For objects with H$\alpha$ detection, the observed H$\alpha$ flux roughly agrees with the expected flux, without significant systematics, though there exists a large scatter of $\sigma\sim0.3$ dex. Although it still remains unclear, the reason of the low detection rate may be due to the combination of various uncertainties when we construct the target sample. Especially, the large uncertainty of the expected H$\alpha$ flux may cause the low detection rate in our survey. The detailed discussions on the H$\alpha$ detection rate is beyond the scope of this work and will be investigated in the future. The obtained spec-$z$ ranges from $z=1.2$ to $z=1.6$ with a median of $z=1.42$. Examples of the obtained spectra with the position of the OH-masks are presented in Figure 2 of Y12, in which spectra with [N\,{\sc ii}]$\lambda$6584 detection with S/N$\ge$ 3.0, 1.5 $\le$ S/N $<$ 3.0, S/N $<$ 1.5 are shown, respectively.

The S/N ratios of [N\,{\sc ii}]$\lambda$6584 line of $\sim$70\% and $\sim$42\% of objects in our sample are $<3.0$ and $<1.5$, respectively. In order to reveal the average metallicity including these low S/N objects, we apply the stacking analysis. As we describe in the following sections, we separate our sample into several groups/bins, stack the individual spectra, and measure the [N\,{\sc ii}]$\lambda$6584/H$\alpha$ line ratio and metallicity from the spectral fitting. In Y12, they perform the stacking analysis in two ways: One is stacking the best-fitting spectra which are corrected properly for the OH-mask effect as described above and in Y12 for more details (\textit{method 1}), and the other is simply stacking the observed spectra (\textit{method 2}). In the latter method, the effects of the OH-mask are not corrected properly. The details of the stacking procedures are explained in Section 3.6 of Y12. In Figure \ref{fig:Spectra}, we show the stacked spectra in five stellar mass bins in the case of both \textit{method 1} and \textit{method 2}. In this paper, as a fiducial stacking method, we use \textit{method 1} unless otherwise noted. 

\begin{figure}
\includegraphics[angle=-90,scale=0.50]{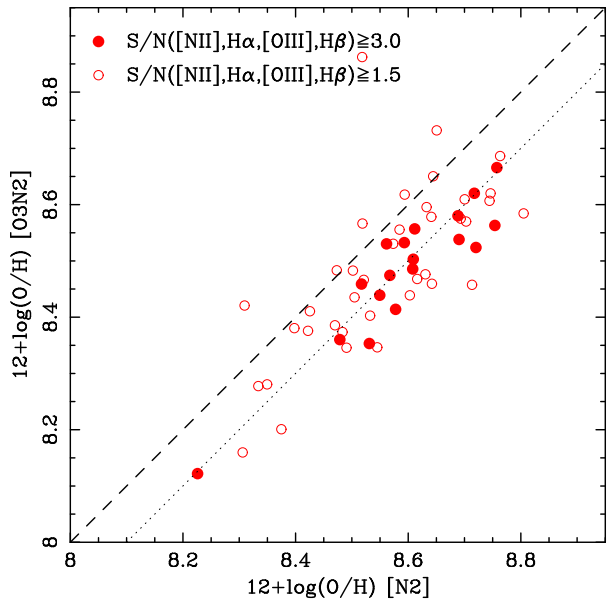}
\caption{Comparison of the metallicity derived from the N2 method and the O3N2 method. Objects with moderate detections (S/N$>$1.5) of all of [N\,{\sc ii}]$\lambda$6584, H$\alpha$, [O\,{\sc iii}]$\lambda$5007, and H$\beta$ lines are indicated by open circles, while significant detections (S/N$>$3.0) are indicated by filled circles. Here, AGN candidates and objects with large mask loss rate are excluded. The dashed line is an equivalent line and the dotted line is that shifted downward by 0.1 dex. \label{fig:O3N2}}
\end{figure}

\begin{figure}
\includegraphics[angle=-90, scale=0.50]{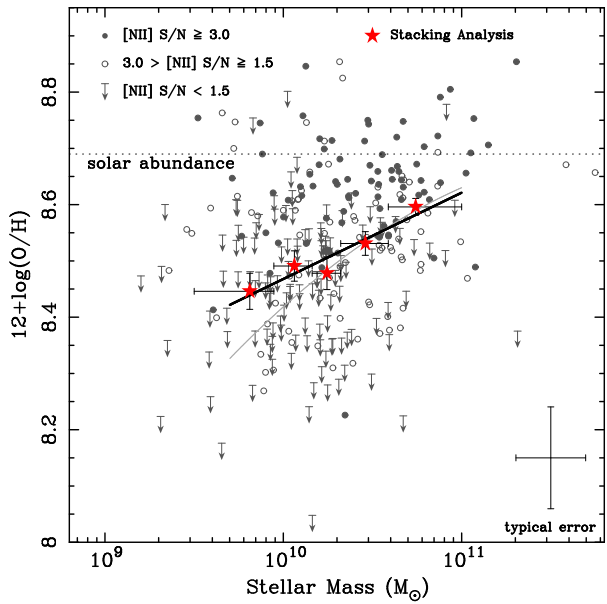}
\caption{Metallicity against stellar mass for our sample. Objects with [N\,{\sc ii}]$\lambda$6584 lines with S/N $>$ 3.0 and 1.5 $<$ S/N $<$ 3.0 are indicated by filled and open symbols, respectively. Those with [N\,{\sc ii}]$\lambda$6584 lines with S/N $<$ 1.5 are plotted as upper limits with values corresponding to $1.5\sigma$. The typical errors of stellar mass and metallicity are shown in the lower right corner. Note that the error of the metallicity is derived from the flux error of H$\alpha$ and [N\,{\sc ii}]$\lambda$6584 lines and does not include the uncertainty of the metallicity calibration. The results of stacking analysis with the bootstrap errors are presented by filled stars. The thick solid line shows the linear fit for the stacking results, while the thin solid line shows the second-order polynomial for the stacking results of Y12. The horizontal dotted line indicates solar metallicity (12+log(O/H)=8.69; \citet{Asplund:2009p11853}).\label{fig:MZR}}
\end{figure}

\begin{figure}
\includegraphics[angle=-90, scale=0.50]{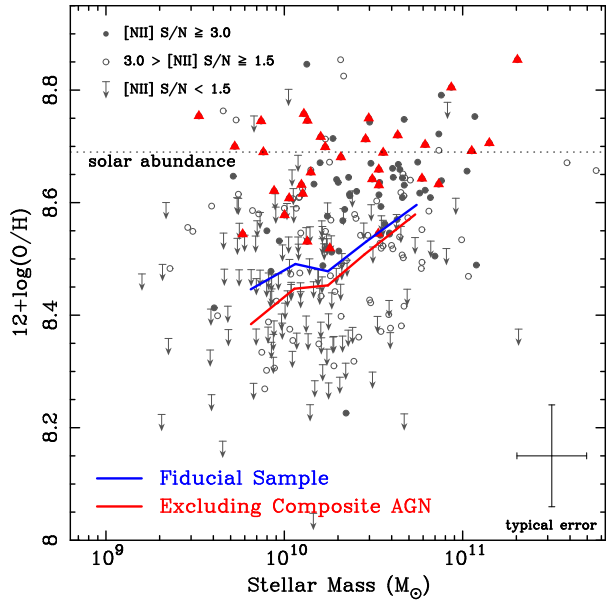}
\caption{The same as Fig. \ref{fig:MZR}. Objects locating in the composite region of the BPT diagram (left panel of Fig. \ref{fig:BPT}) are indicated by red filled triangles. While our fiducial result of the stacking analysis is indicated by the blue solid line, the result excluding the composite objects is indicated by the red solid line. The horizontal dotted line indicates solar metallicity.\label{fig:MZR_ExComp}}
\end{figure}

\section[]{Results and Discussions\label{sec:results}}
\subsection{Optical AGN Diagnostics\label{sec:AGN}}
Although we excluded the X-ray luminous AGN from our sample (Section \ref{sec:PrimarySample}), it is possible that the sample is contaminated by obscured AGNs. Since our observations with FMOS in LR mode cover the wavelength region of $\sim1.0-1.7$ $\mu$m ($\sim4000-7000$ \textrm{\AA} in the rest-frame for the typical redshift of our sample), the line ratio diagnostics by using [N\,{\sc ii}]$\lambda$6584/H$\alpha$ and [O\,{\sc iii}]$\lambda$5007/H$\beta$ line ratios \citep[][hereafter BPT]{Baldwin:1981p15095} can be used to separate the AGN and the normal SFGs. Among the H$\alpha$ detected sample ($\sim343$ objects), [N\,{\sc ii}]$\lambda$6584, H$\alpha$, [O\,{\sc iii}]$\lambda$5007, and H$\beta$ lines are detected with S/N$\ge$2.5 for 33 objects (9.6\%), [N\,{\sc ii}]$\lambda$6584, H$\alpha$, and [O\,{\sc iii}]$\lambda$5007 lines are detected with S/N$\ge$2.5 but H$\beta$ lines are not detected for 55 objects (16.0\%), [N\,{\sc ii}]$\lambda$6584, H$\alpha$, and H$\beta$ lines are detected with S/N$\ge$2.5 but [O\,{\sc iii}]$\lambda$5007 lines are not detected for 33 objects (9.6\%). Although here we use the S/N of 2.5 as the line detection in order to expand the sample for the discussions on the AGN contamination, the following results do not change largely if we adopt the S/N threshold of 3.0.

In the left panel of Figure \ref{fig:BPT}, the distribution of our sample on the BPT diagram is plotted with the empirical criterion to separate the AGN and the SFG by \citet{Kauffmann:2003p8323} and the maximum theoretical line of starburst galaxies by \citet{Kewley:2001p5158}. 15 objects of our sample are located in the AGN region of the theoretical prediction by \citet{Kewley:2001p5158} and these objects are taken as the AGN candidates (indicated by blue squares in Figure \ref{fig:BPT}). The AGN fraction determined from the BPT diagram is 12.4\%, which agrees roughly the AGN fraction among galaxies with stellar mass larger than $10^{9.5}$ M$_{\sun}$ at $1.0<z<2.0$ \citep{Xue:2010p25909}; the fraction of AGNs with $L_{X}=10^{41.9-43.7}$ erg s$^{-1}$ is $\sim 10 \%$. In addition to the BPT AGN candidates, we exclude 3 objects with the line FWHM larger than 1000 km s$^{-1}$ and 3 objects with significantly high [N\,{\sc ii}]$\lambda$6584/H$\alpha$ ratio (log([N\,{\sc ii}]$\lambda$6584/H$\alpha$)$>$0.1, i.e., 12+log(O/H)$>$8.96). In Figure \ref{fig:BPT}, the stacking results presented with filled (for all sample but excluding the AGN candidates) and open (for three sub-samples grouped by stellar mass) stars show that the our sample is not contaminated by the AGN in general. Figure \ref{fig:BPT} also shows that a considerable number of our sample is located between the theoretical line by \citet{Kewley:2001p5158} and the empirical line, often referred to as a composite region; similar trends were reported at $z\sim2$ \citep[e.g.,][]{Erb:2006p4143, Hainline:2009p7444}. In our sample, 43 objects are in the composite region of the BPT diagram (indicated by green circles in Figure \ref{fig:BPT}). The line separating AGNs and SFGs is claimed to evolve with redshift. In Figure \ref{fig:BPT}, we also show the proposed line at $z\sim1.4$ by \citet{Kewley:2013p25967} based on the recent theoretical model by \citet{Kewley:2013p25966}. A large part of objects in the composite region are located within the new theoretical line, which may indicate that most of our sample galaxies are dominated by purely star-forming galaxies.

As we mentioned in Section \ref{sec:SpectralFitting}, the H$\alpha$ and H$\beta$ absorption may affect the emission line ratio. The true H$\alpha$ and H$\beta$ may be larger than actually derived due to this effect and the position of objects on the BPT diagram may change systematically. As we mentioned in Section \ref{sec:SpectralFitting}, since the contribution by the absorption to the individual objects is uncertain, we do not apply the absorption correction for the H$\alpha$ and H$\beta$ emission line, and here we only consider the average effect on the various results. In the case of the maximum stellar absorption, the maximum contribution of the absorption to the emission line is estimated to be $\sim5\%$ and $\sim20\%$ at the flux limit of H$\alpha$ and H$\beta$, respectively. If we consider the effect of stellar absorption, both [N\,{\sc ii}]$\lambda6584$/H$\alpha$ and [O\,{\sc iii}]$\lambda$5007/H$\beta$ line ratios (filled star in Figure \ref{fig:BPT}) decrease systematically by 0.01 dex and 0.09 dex, respectively. Only 4 objects identified as AGNs without the consideration of the absorption effect are identified as non-AGNs if we consider the absorption effect. The consequent effect on metallicity, which is described in the subsequent sections, is $\le 0.01$ dex.

The diagnostic for the SFG and AGN separation by using the stellar mass and the [O\,{\sc iii}]$\lambda$5007/H$\beta$ line ratio is recently proposed. \citet{Juneau:2011p17996} presented that AGNs and SFGs are well separated on the stellar mass - [O\,{\sc iii}]$\lambda$5007/H$\beta$ diagram (hereafter MEx diagram) up to $z\sim1$. The right panel of Figure \ref{fig:BPT} shows our sample on the MEx diagram with the criterion proposed by \citet{Juneau:2011p17996}. It is shown that the possible AGN candidates selected from the BPT diagram are located mainly in the AGN region in the MEx diagram. It is also shown that a part of our sample which is in the SF region in the BPT diagram are located in the AGN region in the MEx diagram. Some of the SF galaxies in the AGN region of the MEx diagram may be affected by the H$\beta$ absorption as we mentioned above. Since the AGN/SFG separation by using the MEx diagram is not calibrated well at $z\ga1$, we do not use the MEx diagnostics for our AGN rejection.

In summary, we exclude 21 objects as AGN candidates from our sample, and define remaining 322 galaxies as the SF galaxy sample. We use this sample for various discussions in the subsequent sections unless otherwise noted.

\begin{figure}
\includegraphics[angle=-90, scale=0.50]{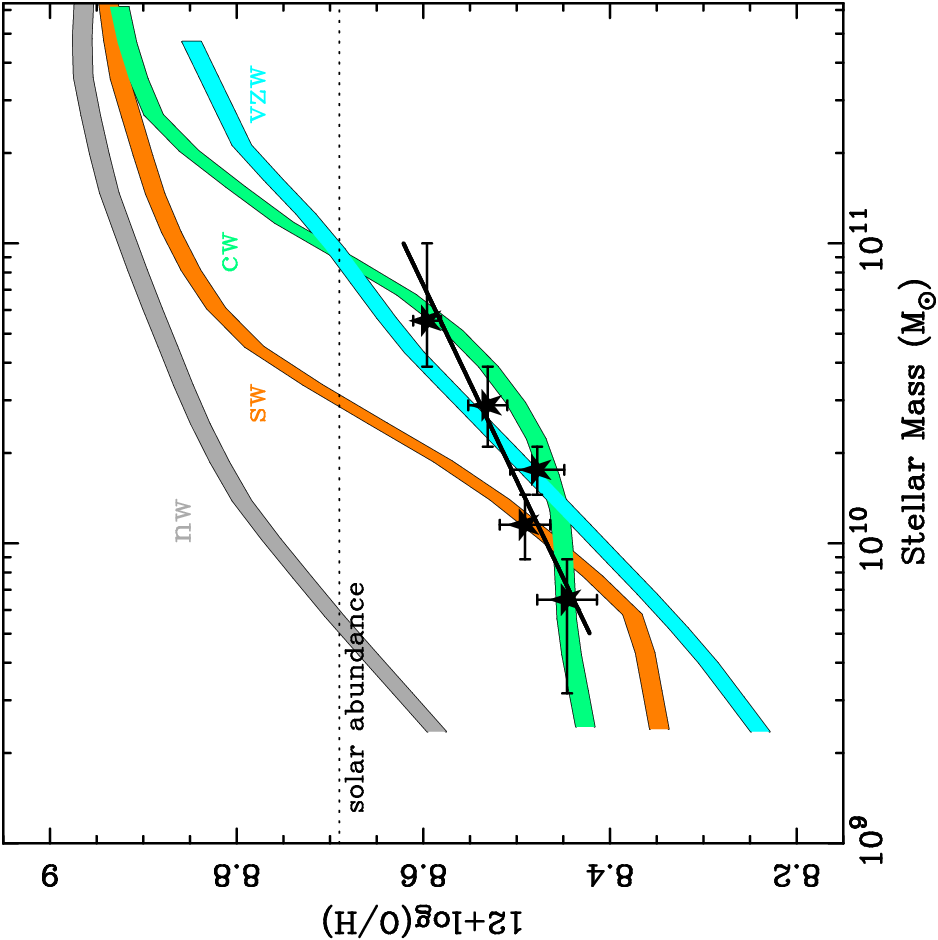}
\caption{Comparison to the theoretical predictions based on the cosmological simulations by \citet{Dave:2011p11823}. The results of our observations at $z\sim1.4$ are plotted by filled stars with error bars. The theoretical models are obtained by averaging the results at $z=1.0$ and $z=2.0$ presented by \citet{Dave:2011p11823}. No winds (nw), Constant winds (cw), Slow winds (sw), and Momentum-conserving winds (vzw) are indicated by gray, green, orange, and cyan, respectively. \label{fig:MZR_Comparison_Models}}
\end{figure}

\begin{table*}
 \begin{center}
 \caption{The median stellar mass, the number of galaxies, metallicity from the stacked spectra, the median SFR from H$\alpha$ luminosity corrected for dust extinction in five stellar mass bins.\label{Tab1}}
 \begin{tabular}{cccc}
 \hline
log($M_{*}$/M$_{\sun}$) & number of galaxies & 12+log(O/H) & SFR (M$_{\sun}$ yr$^{-1}$) \\
\hline
  9.81$_{-0.31}^{+0.13}$ & 55 & 8.45$\pm$0.03 & 39.8$\pm$3.7 \\
10.06$_{-0.12}^{+0.10}$ & 54 & 8.49$\pm$0.03 & 42.9$\pm$3.6 \\
10.25$_{-0.09}^{+0.08}$ & 54 & 8.48$\pm$0.03 & 60.1$\pm$7.5 \\
10.46$_{-0.13}^{+0.13}$ & 54 & 8.53$\pm$0.02 & 73.9$\pm$6.2 \\
10.74$_{-0.15}^{+0.26}$ & 54 & 8.60$\pm$0.02 & 79.4$\pm$8.6 \\
\hline
\end{tabular}
\end{center}
\end{table*}

\begin{table*}
 \begin{center}
 \caption{The median stellar mass, the number of galaxies, metallicity from the stacked spectra, the median SFR from H$\alpha$ luminosity corrected for dust extinction in three stellar mass bins.\label{Tab2}}
 \begin{tabular}{cccc}
 \hline
log($M_{*}$/M$_{\sun}$) & number of galaxies & 12+log(O/H) & SFR (M$_{\sun}$ yr$^{-1}$)\\
\hline
  9.91$_{-0.41}^{+0.18}$ & 91 & 8.44$\pm$0.02 & 39.8$\pm$2.8 \\
10.25$_{-0.16}^{+0.14}$ & 90 & 8.49$\pm$0.02 & 57.9$\pm$5.5 \\
10.66$_{-0.27}^{+0.34}$ & 90 & 8.58$\pm$0.01 & 84.3$\pm$5.8 \\
\hline
\end{tabular}
\end{center}
\end{table*}

\subsection{The Mass-Metallicity Relation\label{sec:MZR}}
The metallicity of our sample galaxy is determined from the N2 ($\equiv$ log([N\,{\sc ii}]$\lambda$6584/H$\alpha$)) index by using the empirical metallicity calibration by \citet{Pettini:2004p7356} (hereafter N2 method):

\begin{equation}
12+\textrm{log(O/H)}=8.90+0.57\times\textrm{N2}.
\end{equation}
The scatter of the calibration itself is $\sim$0.18 dex at 1$\sigma$ significance. It is known that [N\,{\sc ii}]$\lambda$6584 emission lines tends to be only weakly sensitive to metallicity near and above solar metallicity. In order to avoid the saturation effect of the metallicity calibration, \citet{Pettini:2004p7356} present a calibration by using the O3N2 ($\equiv$ log\{([O\,{\sc iii}]$\lambda$5007/H$\beta$)/([N\,{\sc ii}]$\lambda6584$/H$\alpha$)\}) index (hereafter O3N2 method):

\begin{equation}
12+\textrm{log(O/H)}=8.73-0.32\times\textrm{O3N2}.
\end{equation}

About 20 \% of our sample have moderate detections (S/N$\ge$1.5) of all of [N\,{\sc ii}]$\lambda$6584, H$\alpha$, [O\,{\sc iii}]$\lambda$5007, and H$\beta$ emission lines. For these objects, the metallicities measured by the O3N2 method are compared to those by the N2 method in Figure \ref{fig:O3N2}. The obtained metallicity derived by using the O3N2 method is systematically smaller than that from the N2 method by $\sim$0.1 dex. As we mentioned in Section \ref{sec:SpectralFitting}, the both indices could be affected by the Balmer absorption lines. If we assume the maximum absorption case, metallicity from the O3N2 index increases by $\sim0.03$ dex, while metallicity from the N2 index decrease by only $\sim0.005$ dex. However, even if we consider the maximum stellar absorption effect, the O3N2 index is still systematically smaller than the N2 index, and therefore we conclude that the difference is real. These objects with large offsets between the O3N2 and the N2 index show larger [O\,{\sc iii}]$\lambda$5007/H$\beta$ and/or [N\,{\sc ii}]$\lambda$6584/H$\alpha$ line ratios with respect to the sequence of local SFGs in the BPT diagram; a part of them is located in the composite region of the BPT diagram in Figure \ref{fig:BPT}. This could be due to the different physical conditions, for instance, high ionization parameters \citep{Kewley:2002p5161,Erb:2006p4143,Liu:2008p5650,Kewley:2013p25967}, in SFGs at high redshift as compared to local ones. In this work, we use the N2 method as a fiducial way to measure the metallicity of our sample.

Figure \ref{fig:MZR} shows the distribution of the metallicity against the stellar mass for our sample at $z\sim1.4$. Objects showing [N\,{\sc ii}]$\lambda$6584 emission lines with S/N $>$ 3.0 and 1.5 $<$ S/N $<$ 3.0 are indicated by filled and open circles, respectively. For those showing [N\,{\sc ii}]$\lambda$6584 emission lines with S/N $<$ 1.5, we take the values corresponding to $1.5\sigma$ as upper limits. The stellar masses are recalculated from the SED fitting with the redshift fixed to the observed spec-$z$ and we use the stellar masses hereafter; the difference from the original stellar masses is very small as we discussed in Section \ref{sec:PrimarySample}. 

It is found that the more massive galaxies tend to have the higher metallicity, though there exists considerable scatter larger than the observational errors. Since the obtained metallicities of many objects are actually upper limits, in order to obtain the average metallicity, our sample is divided into several stellar mass bins and the obtained spectra are stacked in each stellar mass bin including the upper limit objects. As we described in Section \ref{sec:SpectralFitting}, we carried out the stacking analysis in two ways (\textit{method 1} and \textit{method 2}), though we take the results derived from \textit{method 1} as our fiducial results throughout this paper. Firstly our sample is divided into five stellar mass bins, and the metallicity is derived from the stacked spectrum in each stellar mass bin by using the N2 method. The resultant metallicities from the stacking analysis are 12$+$log(O/H)$=$8.45, 8.49, 8.48, 8.53, and 8.60 in the stellar mass bin of log($M_{*}$/M$_{\sun}$)$=$9.81,10.06, 10.25, 10.46, and 10.74, respectively. The stacked spectra are also presented in Figure \ref{fig:Spectra}. If the sample is divided into 3 mass bins, the stacking results are 12$+$log(O/H)$=$8.44, 8.49, and 8.58 in the stellar mass bin of log($M_{*}$/M$_{\sun}$)$=$9.91,10.25, and 10.66, respectively. The results are also summarized in Table \ref{Tab1} and Table \ref{Tab2}. The errors are estimated based on the bootstrap resampling method by running 1000 trials. Although the resulting metallicity with \textit{method 2} is comparable to that obtained with \textit{method 1} in the massive part, the metallicity with \textit{method 2} is up to $\sim0.07$ dex smaller than that by \textit{method 1} in the low mass part. In this paper, again, we use results from \textit{method 1} as the fiducial results.

In Figure \ref{fig:MZR}, the result of the stacking analysis shows the clear mass-metallicity relation; the massive galaxies tend to have higher metallicity, which has been reported with a smaller sample at this redshift by Y12. This trend can also be found in Figure \ref{fig:Spectra}. From the simple least squares fit, the result can be expressed by a linear function:

\begin{equation}
12+\textrm{log(O/H)}=6.93+0.153\times \textrm{log(}M_{*}/\textrm{M}_{\sun}),
\end{equation}
which is shown by the thick solid line in Figure \ref{fig:MZR}, while the polynomial fit presented by Y12 is presented by thin solid line. Although the two results agree with each other within the error bars, the mass-metallicity relation by Y12 is somewhat under-estimated in the low mass stellar mass bins probably due to the small size of the sample by Y12. The resulting mass-metallicity relation at $z\sim1.4$ is comparable to that derived from the spectral stacking with 11 galaxies by \citet{Liu:2008p5650} within $\pm0.02$ dex. We note that they use the same metallicity calibration as ours, i.e, N2 method, but they assume the Chabrier IMF, therefore we convert their stellar mass to that with the Salpeter IMF.

As mentioned in Section \ref{sec:AGN}, our sample includes objects in the composite region on the BPT diagram. The effect of these composite objects on the mass-metallicity relation is examined by excluding the objects from our sample. Figure \ref{fig:MZR_ExComp} shows that the resulting mass-metallicity relation without the composite objects is shifted downward; the metallicity decreases by up to $\sim0.05$ dex at fixed stellar mass if the composite objects are excluded from the sample. Since objects in the composite region generally show larger [N\,{\sc ii}]$\lambda$6584/H$\alpha$ flux ratio as shown in the left panel of Figure \ref{fig:BPT}, the downward shift in the resulting mass-metallicity relation is a reasonable result.

The observed mass-metallicity relation of our sample at $z\sim1.4$ is compared with the theoretical prediction of the cosmological simulations by \citet{Dave:2011p11823} with various wind models: no winds (\textit{nw}), slow winds (\textit{sw}), constant winds (\textit{cw}), and momentum-conserving winds (\textit{vzw}). In Figure \ref{fig:MZR_Comparison_Models}, those of the simulations with the \textit{cw} or the \textit{vzw} wind models can reproduce our result, which implies the necessity for the moderately strong galactic winds. The presence of the strong winds in galaxies at $z\sim1-2$ is reported by other spectroscopic studies \citep{Weiner:2009p13953,Steidel:2010p10312,Newman:2012p23470}.

\begin{figure}
\includegraphics[angle=-90, scale=0.50]{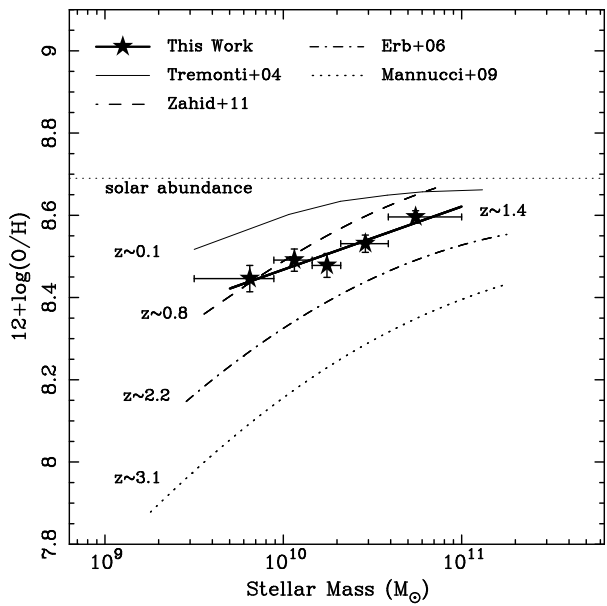}
\caption{Comparison of the mass-metallicity relation to the previous studies at $z\sim0.1$ \citep[thin solid; ][]{Tremonti:2004p4119}, $z\sim0.8$ \citep[dashed; ][]{Zahid:2011p11939}, $z\sim2.2$ \citep[dashed-dotted; ][]{Erb:2006p4143}, and $z\sim3.1$ \citep[dotted; ][]{Mannucci:2009p8028}. For each line, the stellar mass range actually observed is presented. Both the stellar mass and metallicity of other samples are converted so that the IMF and metallicity calibration are consistent with those we adopted. The horizontal dotted line indicates solar metallicity.\label{fig:MZR_Comparison}}
\end{figure}

\begin{figure}
\includegraphics[angle=-90, scale=0.50]{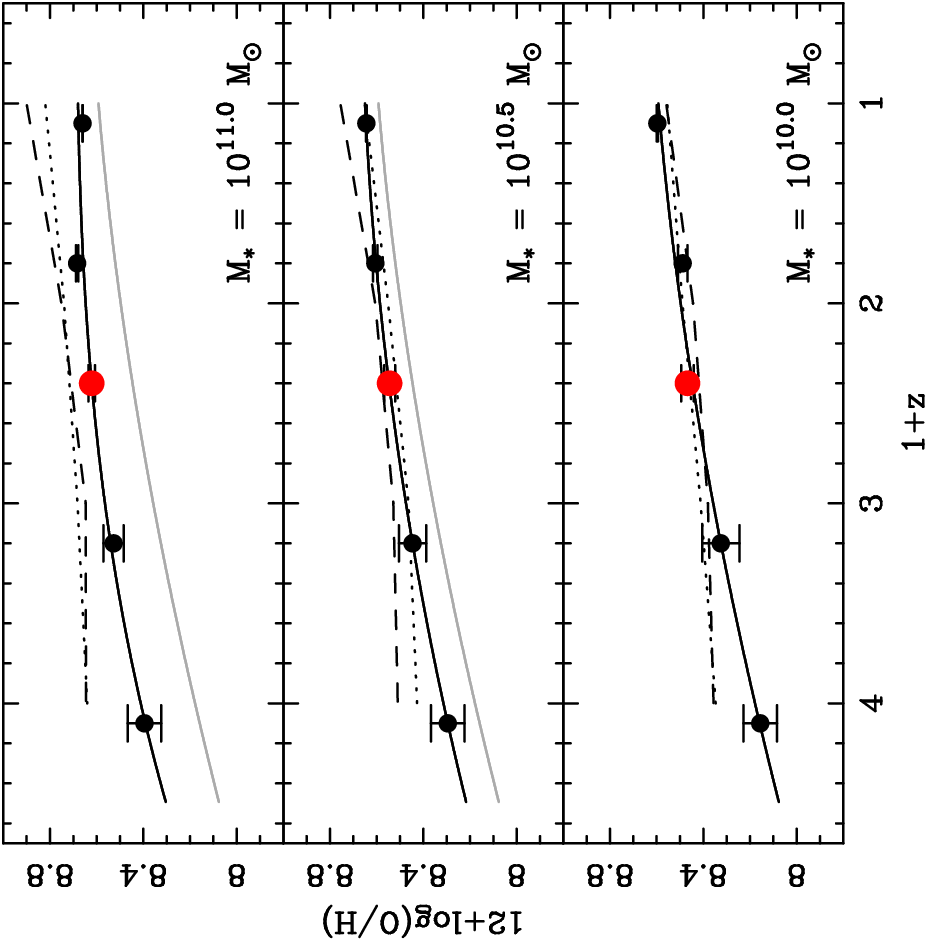}
\caption{Cosmic evolution of the mean metallicity at $M_{*}=10^{10.0}$ M$_{\sun}$ (bottom), $10^{10.5}$ M$_{\sun}$ (middle), and $10^{11.0}$ M$_{\sun}$ (top). The data points at $z\sim0.1$ \citep{Tremonti:2004p4119}, $z\sim0.8$ \citep{Zahid:2011p11939}, $z\sim1.4$ (this work), $z\sim2.2$ \citep{Erb:2006p4143}, and $z\sim3.1$ \citep{Mannucci:2009p8028} are plotted. In each panel, the black solid line shows the best-fitting as a function of $1+z$. The gray solid line in top two panel is the best-fitting function at $M_{*}=10^{10.0}$ M$_{\sun}$. Dashed and dotted lines show the \textit{vzw} and \textit{cw} models from the cosmological simulations by \citet{Dave:2011p11823}, respectively. \label{fig:CME}}
\end{figure}

\subsection{Cosmic Evolution of the Mass-Metallicity Relation\label{sec:MZR_Evolution}}

The cosmic evolution of the mass-metallicity relation is argued in many works \citep[e.g.,][]{Savaglio:2005p3325,Erb:2006p4143,Maiolino:2008p5212}. In Figure \ref{fig:MZR_Comparison}, the mess-metallicity relation at $z\sim1.4$ is compared with previous results at $z\sim0.1$ \citep[][; here we use the recalculated results with the N2 indicator by \citet{Erb:2006p4143}.]{Tremonti:2004p4119}, $z\sim0.8$ \citep{Zahid:2011p11939}, $z\sim2.2$ \citep{Erb:2006p4143}, and $z\sim3.1$ \citep{Mannucci:2009p8028}. In order to make a fair comparison, the conversion from the Chabrier IMF to the Salpeter IMF, which is that we assumed, is applied for the stellar mass of these samples. The metallicity is also converted so that the metallicity calibration is consistent with that we use, i.e., the N2 method, by assuming the conversions by \citet{Kewley:2008p5141} for the sample by \citet{Zahid:2011p11939} and \citet{Nagao:2006p8780} for the sample by \citet{Mannucci:2009p8028}. Figure \ref{fig:MZR_Comparison} shows that the results of our sample at $z\sim1.4$ are located between those at $z\sim0.8$ and $z\sim2.2$, and overall trend of increasing metallicity with cosmic time from $z\sim3.1$ to $z\sim0.1$ can be seen. Our result at $z\sim1.4$, however, is comparable to that at $z\sim0.8$ by \citet{Zahid:2011p11939} at $M_{*}\la10^{10}$ M$_{\sun}$.

In Figure \ref{fig:CME}, the averaged metallicity at the stellar mass of $10^{10.0}$, $10^{10.5}$, and $10^{11.0}$ M$_{\sun}$ is plotted against the redshift from $z\sim3.1$ to $z\sim0.1$. It is clearly shown that the mean metallicity increases with decreasing redshift at any stellar mass. The smooth cosmological evolution of the average metallicity is well reproduced with the function of:
\begin{equation}
\textrm{12+log(O/H)}=8.63-0.041(1+z)^{1.74},
\end{equation}
at $M_{*}=10^{10.0}$ M$_{\sun}$,
\begin{equation}
\textrm{12+log(O/H)}=8.67-0.020(1+z)^{2.08},
\end{equation}
at $M_{*}=10^{10.5}$ M$_{\sun}$,
\begin{equation}
\textrm{12+log(O/H)}=8.69-0.005(1+z)^{2.85},
\end{equation}
at $M_{*}=10^{11.0}$ M$_{\sun}$. In Figure \ref{fig:CME}, the slope of the evolutionary function in the low mass part appears to be steeper that that at massive part; the increase of metallicity from $z\sim3.1$ to $z\sim0.1$ are 0.44, 0.35, and 0.26 dex at $M_{*}=10^{10}$ M$_{\sun}$, $10^{10.5}$ M$_{\sun}$, and $10^{11}$ M$_{\sun}$, respectively. Although we should pay close attention to various systematic uncertainty of the metallicity, the result indicate the mass-dependent evolution of the mass-metallicity relation.

\begin{figure}
\includegraphics[angle=-90, scale=0.50]{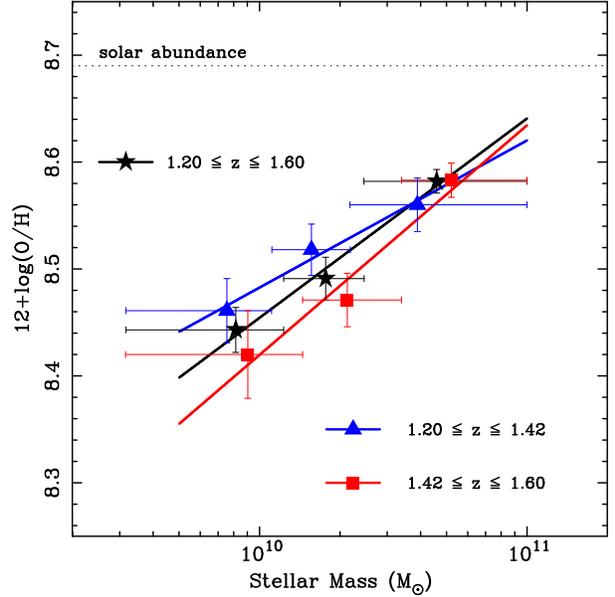}
\caption{The mass-metallicity relation of our sample at $1.20 \le z \le 1.60$ with a median of $z=1.42$ (black stars) and the sub-samples at $1.20 \le z < 1.42$ with a median of $z=1.34$ (blue triangles) and $1.42\le z \le 1.60$ with a median of $z=1.46$ (red squares) are presented. The regression lines for the results of each sample are indicated by solid lines. The horizontal dotted line indicates solar metallicity. \label{fig:MZR_zevol}}
\end{figure}

Although there exists a smooth evolution of the mass-metallicity relation, it is also worth noting that our sample selection is different from that at other redshifts. For a fair comparison of the mass-metallicity relation at different redshifts, it is desirable to compare samples that are selected by the same selection method and the metallicities are derived by using the same calibration. In order to see the evolution fairly, we divide our sample into two groups according to their spec-$z$s ($1.20<z<1.42$ with median $z=1.34$ and $1.42<z<1.60$ with median $z=1.46$) and three stellar mass bins. For each group, the mass-metallicity relation is derived from the stacking analysis by using \textit{method 1} described in Section \ref{sec:SpectralFitting}. Although the error bars of the obtained metallicity from stacked spectra are relatively large, the result presented in Figure \ref{fig:MZR_zevol} may imply the evolution of the mass-metallicity relation in the narrow redshift range ($\sim0.32$ Gyr from $z\sim1.46$ to $z\sim1.34$). It is interesting that low mass galaxies show a stronger metallicity evolution than more massive galaxies over the redshift range of $z\sim1.3$ to $z\sim1.5$, which is consistent with the overall trend of the mass-metallicity relation described above. Although the saturation effects of the N2 indicator may affect the results at the massive part, the metallicity at the most massive bins are below the solar abundance where the N2 calibration is still robust. Since we made the expected H$\alpha$ flux cut in our sample selection, the high-$z$ sample at $z\sim1.5$ would be biased towards higher SFR. This may also cause a bias to the metallicity. The difference of the intrinsic SFR between two samples is small: The average SFRs of the sample at $z\sim1.3$ and $z\sim1.5$ are 60 M$_{\sun}$ yr$^{-1}$ and 64 M$_{\sun}$ yr$^{-1}$, respectively. The effect of the differing SFR on the metallicity is also small at this redshift and SFR range, as we described in Section \ref{sec:MZR_dependence} in details.

In Figure \ref{fig:CME}, we compare our observational results to theoretical predictions of the evolution of metallicity at fixed stellar mass. We show the \textit{cw} and \textit{vzw} models from \citet{Dave:2011p11823} which provide the best fit to our mass-metallicity relation at $z\sim1.4$ (Figure \ref{fig:MZR_Comparison_Models}). These models include galactic winds, but with different prescriptions for mass outflow rates. Both models are consistent with the data at $z\la2$ for the $10^{10.0}$ M$_{\sun}$ and $10^{10.5}$ M$_{\sun}$ mass bins. At $10^{11.0}$ M$_{\sun}$, the models lie above the data by $0.1-0.2$ dex. At $z\sim3$, the models lie above the data by $0.15-0.2$ dex at all stellar masses.

\begin{table*}
 \begin{center}
 \caption{Metallicity Dependence of Various Physical Parameters. For each parameter and stellar mass bin, the threshold and the median values of each sub-group are presented. The details are described in Section \ref{sec:MZR_dependence}.\label{Tab3}}
 \begin{tabular}{clccc}
 \hline
parameter & & mass bin 1 & mass bin 2 & mass bin 3\\
\hline
SFR (H$\alpha$) & Threshold (M$_{\sun}$yr$^{-1}$) & 39.8 & 57.9 & 84.3 \\
& SFR$^{low}$ (M$_{\sun}$yr$^{-1}$)  & 29.6$\pm$1.3 & 43.5$\pm$1.8 & 58.2$\pm$2.3 \\
& SFR$^{high}$ (M$_{\sun}$yr$^{-1}$)  & 61.5$\pm$3.4 & 87.1$\pm$8.3 & 119.2$\pm$7.5 \\
& log($M_{*}^{low}$/M$_{\sun}$) & 9.90$_{-0.31}^{+0.19}$ & 10.23$_{-0.14}^{+0.15}$  & 10.65$_{-0.26}^{+0.31}$ \\
& log($M_{*}^{high}$/M$_{\sun}$) & 9.92$_{-0.40}^{+0.15}$ & 10.26$_{-0.17}^{+0.13}$ & 10.68$_{-0.20}^{+0.32}$ \\
& 12+log(O/H)$^{low}$ & 8.42$\pm$0.04 & 8.56$\pm$0.04 & 8.59$\pm$0.02 \\
& 12+log(O/H)$^{high}$ & 8.45$\pm$0.02 & 8.47$\pm$0.03 & 8.57$\pm$0.01 \\
\hline
SFR (UV) & Threshold (M$_{\sun}$yr$^{-1}$) & 25.0 & 46.9 & 99.6 \\
& SFR$^{low}$ (M$_{\sun}$yr$^{-1}$)  & 20.0$\pm$0.6 & 35.6$\pm$1.0 & 76.0$\pm$2.7 \\
& SFR$^{high}$ (M$_{\sun}$yr$^{-1}$)  & 32.6$\pm$5.0 & 63.7$\pm$3.5 & 146.5$\pm$9.0 \\
& log($M_{*}^{low}$/M$_{\sun}$) & 9.88$_{-0.30}^{+0.20}$ & 10.22$_{-0.13}^{+0.14}$  & 10.60$_{-0.21}^{+0.28}$ \\
& log($M_{*}^{high}$/M$_{\sun}$) & 9.94$_{-0.42}^{+0.15}$ & 10.27$_{-0.18}^{+0.12}$ & 10.72$_{-0.27}^{+0.28}$ \\
& 12+log(O/H)$^{low}$ & 8.43$\pm$0.04 & 8.52$\pm$0.03 & 8.57$\pm$0.02\\
& 12+log(O/H)$^{high}$ & 8.45$\pm$0.03 & 8.47$\pm$0.03 & 8.60$\pm$0.02\\
\hline
sSFR (H$\alpha$) & Threshold (Gyr$^{-1}$) & 5.14 & 3.21 & 2.09\\
& sSFR$^{low}$ (Gyr$^{-1}$)  & 3.65$\pm$0.18 & 2.41$\pm$0.10 & 1.31$\pm$0.07 \\
& sSFR$^{high}$ (Gyr$^{-1}$)  & 8.86$\pm$0.57 & 5.29$\pm$0.42 & 2.87$\pm$0.14 \\
& log($M_{*}^{low}$/M$_{\sun}$) & 9.97$_{-0.30}^{+0.12}$ & 10.26$_{-0.17}^{+0.12}$  & 10.70$_{-0.28}^{+0.27}$ \\
& log($M_{*}^{high}$/M$_{\sun}$) & 9.85$_{-0.32}^{+0.23}$ & 10.23$_{-0.14}^{+0.16}$ & 10.63$_{-0.23}^{+0.37}$ \\
& 12+log(O/H)$^{low}$ & 8.45$\pm$0.03 & 8.54$\pm$0.03 & 8.60$\pm$0.02\\
& 12+log(O/H)$^{high}$ & 8.44$\pm$0.03 & 8.49$\pm$0.03 & 8.57$\pm$0.01\\
\hline
sSFR (UV) & Threshold (Gyr$^{-1}$) & 2.78 & 2.88 & 2.43\\
& sSFR$^{low}$ (Gyr$^{-1}$)  & 2.44$\pm$0.05 & 2.12$\pm$0.07 & 1.74$\pm$0.07 \\
& sSFR$^{high}$ (Gyr$^{-1}$)  & 4.11$\pm$0.99 & 3.46$\pm$0.23 & 3.21$\pm$0.12 \\
& log($M_{*}^{low}$/M$_{\sun}$) & 9.96$_{-0.37}^{+0.13}$ & 10.25$_{-0.16}^{+0.13}$  & 10.66$_{-0.26}^{+0.30}$ \\
& log($M_{*}^{high}$/M$_{\sun}$) & 9.86$_{-0.34}^{+0.19}$ & 10.25$_{-0.16}^{+0.14}$ & 10.66$_{-0.27}^{+0.33}$ \\
& 12+log(O/H)$^{low}$ & 8.44$\pm$0.03 & 8.50$\pm$0.03 & 8.58$\pm$0.02\\
& 12+log(O/H)$^{high}$ & 8.44$\pm$0.03 & 8.48$\pm$0.03 & 8.58$\pm$0.02\\
\hline
$E(B-V)$ & Threshold (mag) & 0.13 & 0.17 & 0.30\\
& $E(B-V)$$^{low}$ (mag)  & 0.10$\pm$0.01 & 0.14$\pm$0.01 & 0.24$\pm$0.01 \\
& $E(B-V)$$^{high}$ (mag)  & 0.17$\pm$0.01 & 0.23$\pm$0.01 & 0.35$\pm$0.01 \\
& log($M_{*}^{low}$/M$_{\sun}$) & 9.90$_{-0.32}^{+0.18}$ & 10.22$_{-0.13}^{+0.17}$  & 10.62$_{-0.22}^{+0.26}$ \\
& log($M_{*}^{high}$/M$_{\sun}$) & 9.92$_{-0.40}^{+0.17}$ & 10.28$_{-0.18}^{+0.11}$ & 10.70$_{-0.31}^{+0.29}$ \\
& 12+log(O/H)$^{low}$ & 8.43$\pm$0.03 & 8.47$\pm$0.03 & 8.56$\pm$0.02\\
& 12+log(O/H)$^{high}$ & 8.46$\pm$0.03 & 8.52$\pm$0.03 & 8.61$\pm$0.02\\
\hline
$B-R$ & Threshold (mag) & 0.24 & 0.32 & 0.56\\
& $B-R$$^{low}$ (mag)  & 0.16$\pm$0.01 & 0.24$\pm$0.01 & 0.43$\pm$0.02 \\
& $B-R$$^{high}$ (mag)  & 0.31$\pm$0.02 & 0.41$\pm$0.02 & 0.68$\pm$0.01 \\
& log($M_{*}^{low}$/M$_{\sun}$) & 9.92$_{-0.33}^{+0.17}$ & 10.22$_{-0.13}^{+0.16}$  & 10.61$_{-0.21}^{+0.27}$ \\
& log($M_{*}^{high}$/M$_{\sun}$) & 9.90$_{-0.38}^{+0.18}$ & 10.27$_{-0.18}^{+0.12}$ & 10.71$_{-0.31}^{+0.29}$ \\
& 12+log(O/H)$^{low}$ & 8.43$\pm$0.04 & 8.46$\pm$0.03 & 8.56$\pm$0.02\\
& 12+log(O/H)$^{high}$ & 8.46$\pm$0.04 & 8.53$\pm$0.03 & 8.60$\pm$0.02\\
\hline
$R-H$ & Threshold (mag) & 1.13 & 1.45 & 2.10\\
& $R-H$$^{low}$ (mag)  & 0.98$\pm$0.02 & 1.26$\pm$0.03 & 1.82$\pm$0.03 \\
& $R-H$$^{high}$ (mag)  & 1.30$\pm$0.03 & 1.66$\pm$0.03 & 2.33$\pm$0.02 \\
& log($M_{*}^{low}$/M$_{\sun}$) & 9.87$_{-0.28}^{+0.22}$ & 10.21$_{-0.12}^{+0.14}$  & 10.58$_{-0.18}^{+0.26}$ \\
& log($M_{*}^{high}$/M$_{\sun}$) & 9.95$_{-0.43}^{+0.14}$ & 10.28$_{-0.19}^{+0.12}$ & 10.73$_{-0.34}^{+0.26}$ \\
& 12+log(O/H)$^{low}$ & 8.42$\pm$0.03 & 8.47$\pm$0.03 & 8.55$\pm$0.02\\
& 12+log(O/H)$^{high}$ & 8.48$\pm$0.03 & 8.52$\pm$0.02 & 8.62$\pm$0.02\\
\hline
$r_{50}$ & Threshold (kpc) & 4.17 & 4.37 & 5.04\\
& $r_{50}$$^{low}$ (kpc)  & 3.63$\pm$0.07 & 3.88$\pm$0.06 & 4.37$\pm$0.07 \\
& $r_{50}$$^{high}$ (kpc)  & 4.67$\pm$0.10 & 5.25$\pm$0.12 & 5.65$\pm$0.06 \\
& log($M_{*}^{low}$/M$_{\sun}$) & 9.91$_{-0.39}^{+0.18}$ & 10.25$_{-0.16}^{+0.14}$  & 10.62$_{-0.22}^{+0.32}$ \\
& log($M_{*}^{high}$/M$_{\sun}$) & 9.91$_{-0.32}^{+0.17}$ & 10.24$_{-0.15}^{+0.14}$ & 10.70$_{-0.31}^{+0.29}$ \\
& 12+log(O/H)$^{low}$ & 8.47$\pm$0.03 & 8.54$\pm$0.02 & 8.58$\pm$0.02\\
& 12+log(O/H)$^{high}$ & 8.40$\pm$0.04 & 8.43$\pm$0.04 & 8.58$\pm$0.01\\
\hline
$f_{gas}$ & Threshold & 0.58 & 0.46 & 0.32\\
& $f_{gas}$$^{low}$ & 0.50$\pm$0.01 & 0.39$\pm$0.01 & 0.25$\pm$0.01 \\
& $f_{gas}$$^{high}$ & 0.67$\pm$0.01 & 0.53$\pm$0.01 & 0.39$\pm$0.01 \\
& log($M_{*}^{low}$/M$_{\sun}$) & 9.97$_{-0.25}^{+0.11}$ & 10.27$_{-0.18}^{+0.11}$  & 10.72$_{-0.30}^{+0.28}$ \\
& log($M_{*}^{high}$/M$_{\sun}$) & 9.83$_{-0.31}^{+0.24}$ & 10.22$_{-0.13}^{+0.17}$ & 10.60$_{-0.20}^{+0.27}$ \\
& 12+log(O/H)$^{low}$ & 8.45$\pm$0.03 & 8.52$\pm$0.03 & 8.57$\pm$0.02\\
& 12+log(O/H)$^{high}$ & 8.44$\pm$0.03 & 8.49$\pm$0.02 & 8.57$\pm$0.01\\
\hline
\end{tabular}
\end{center}
\end{table*}

\begin{figure*}
\includegraphics[angle=-90,scale=0.33]{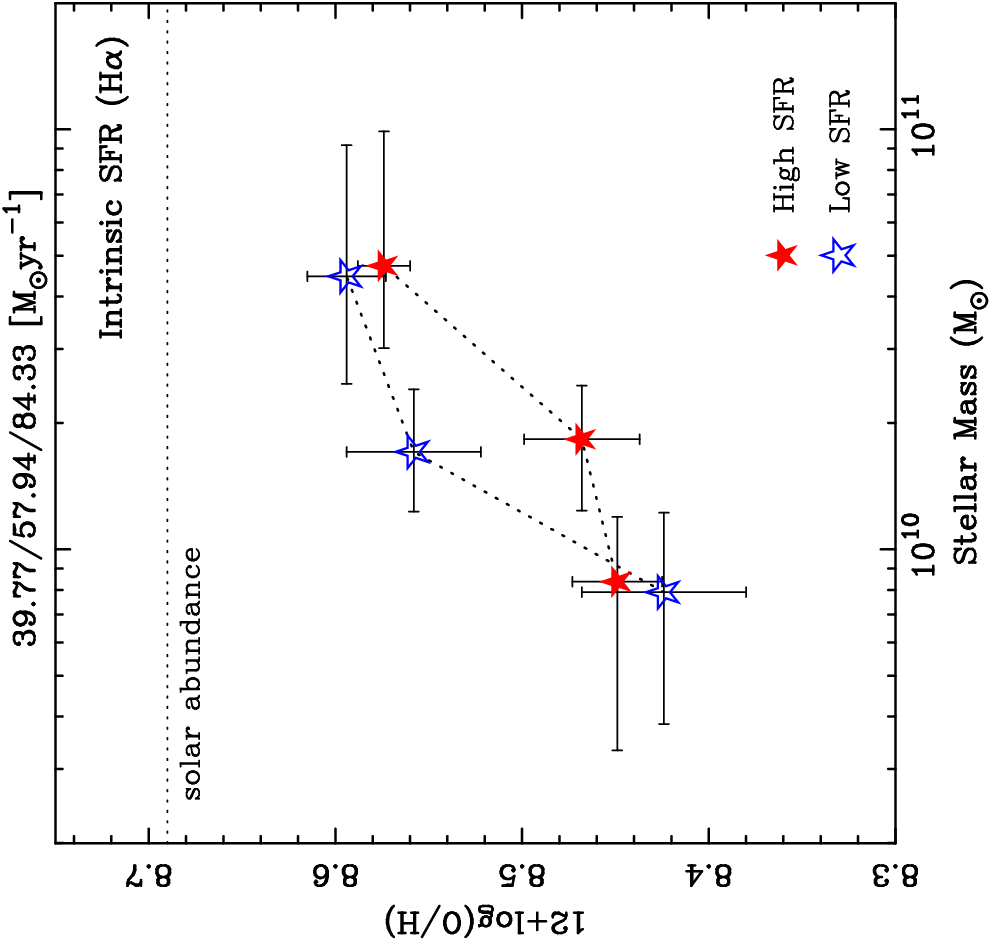}
\includegraphics[angle=-90,scale=0.33]{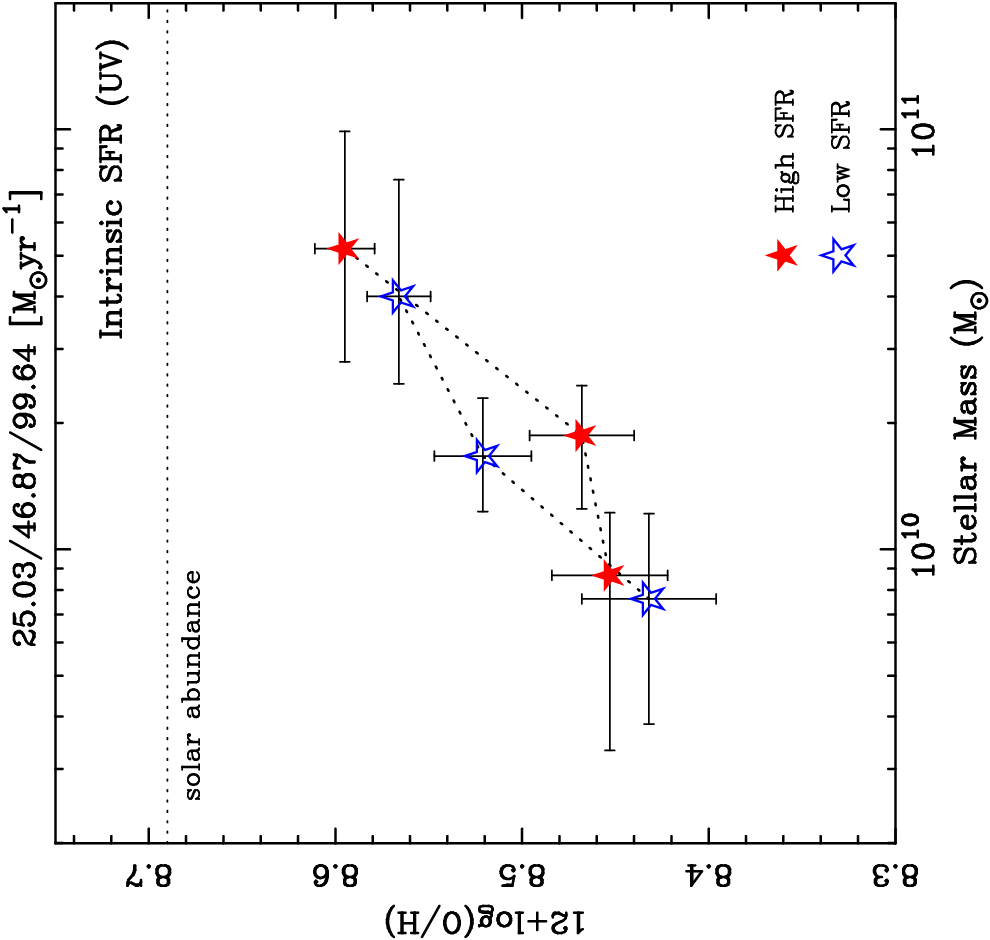}
\includegraphics[angle=-90,scale=0.33]{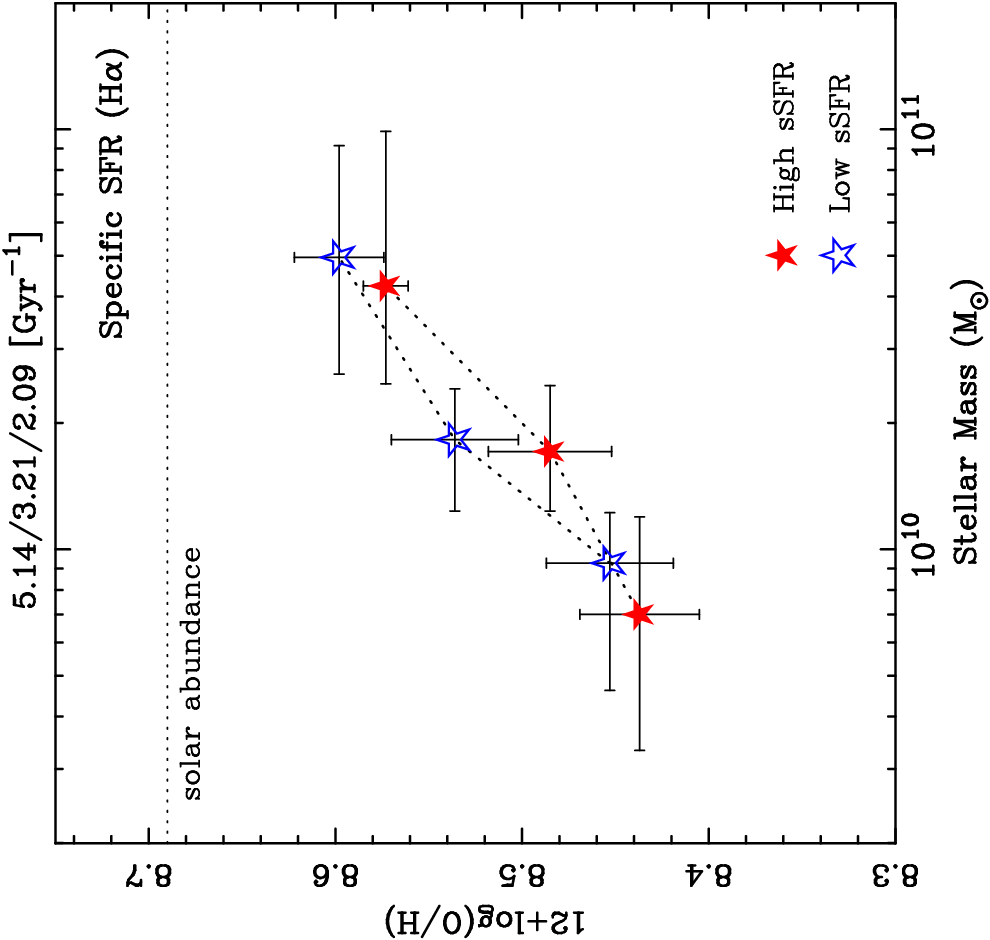}\\
\includegraphics[angle=-90,scale=0.33]{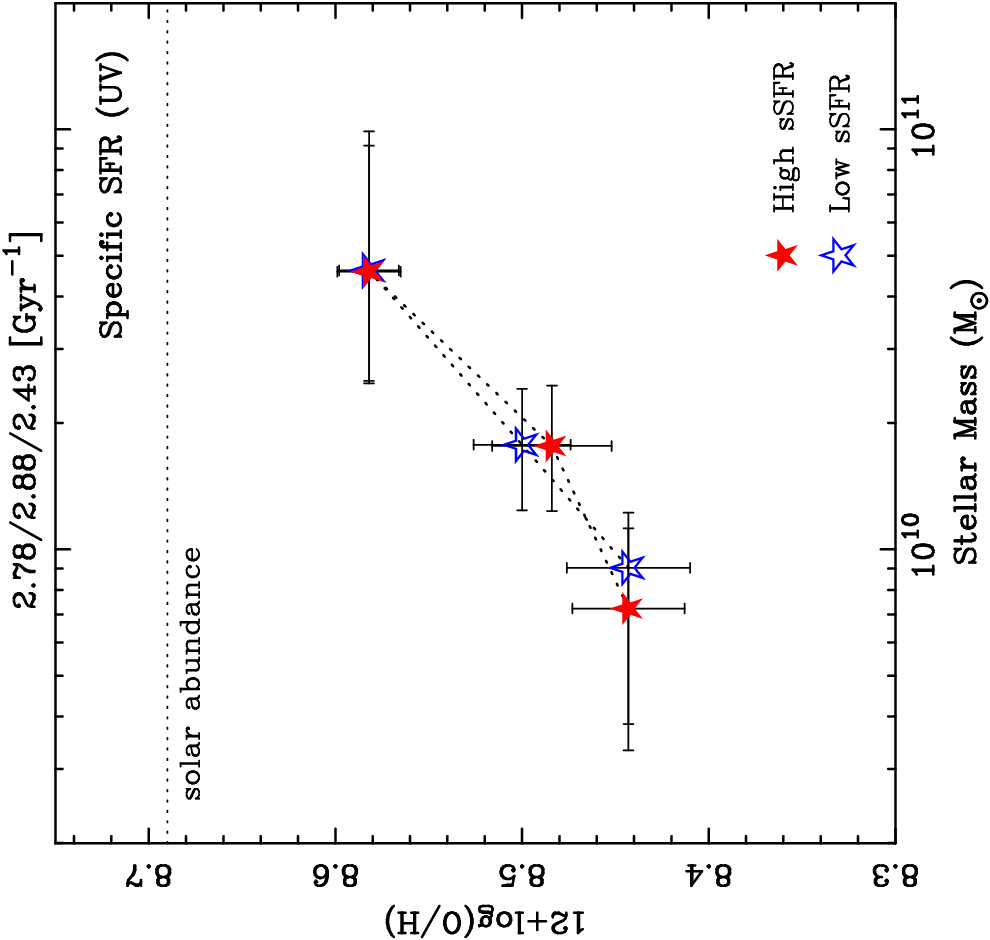}
\includegraphics[angle=-90,scale=0.33]{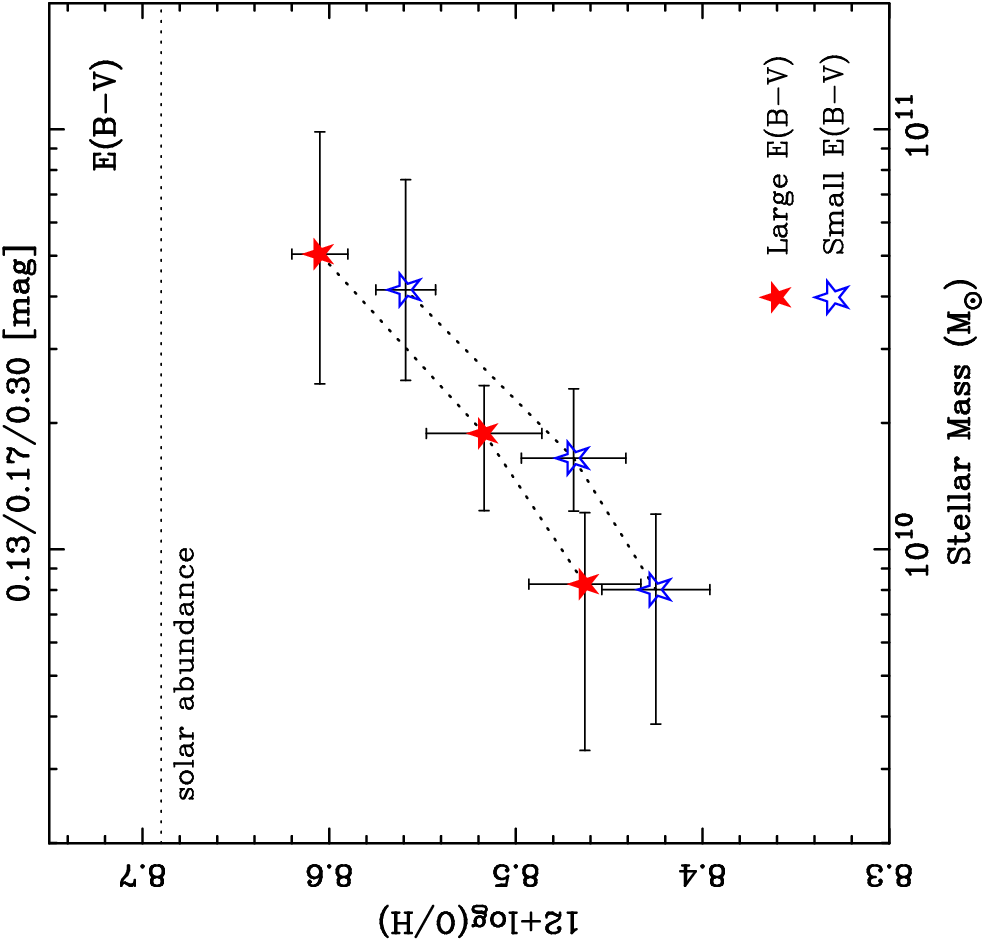}
\includegraphics[angle=-90,scale=0.33]{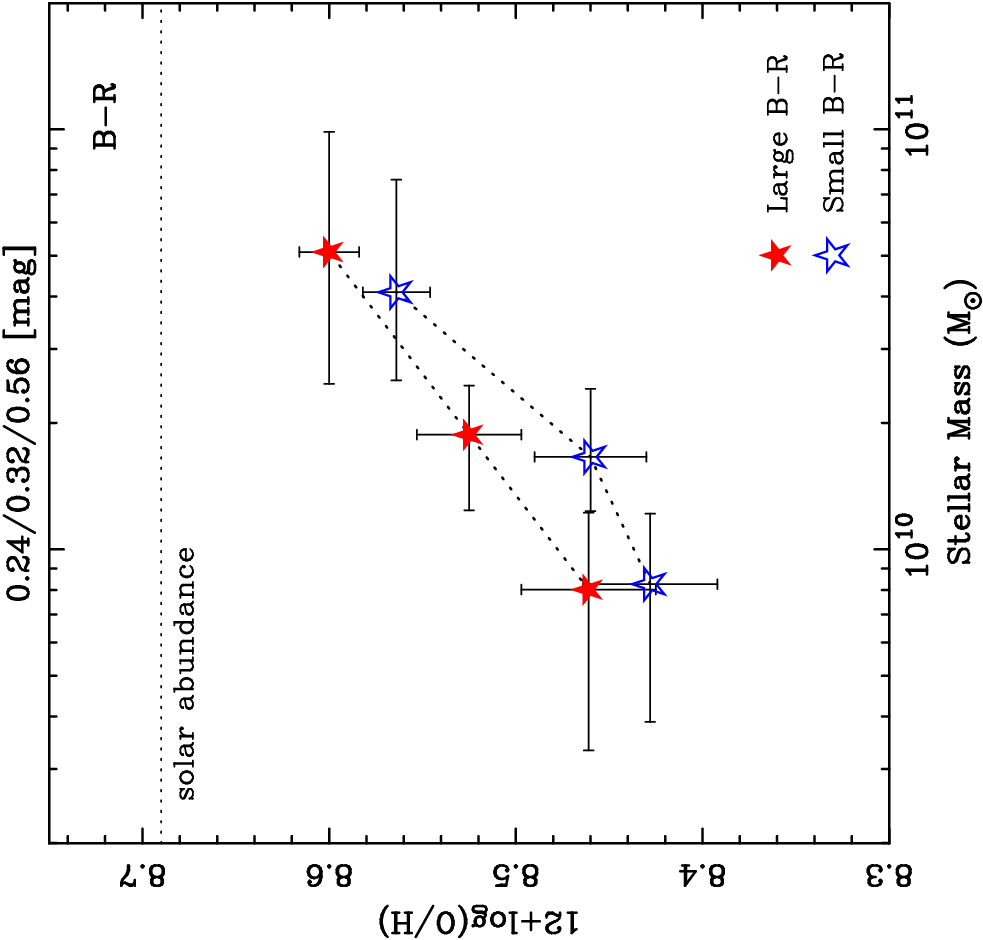}\\
\includegraphics[angle=-90,scale=0.33]{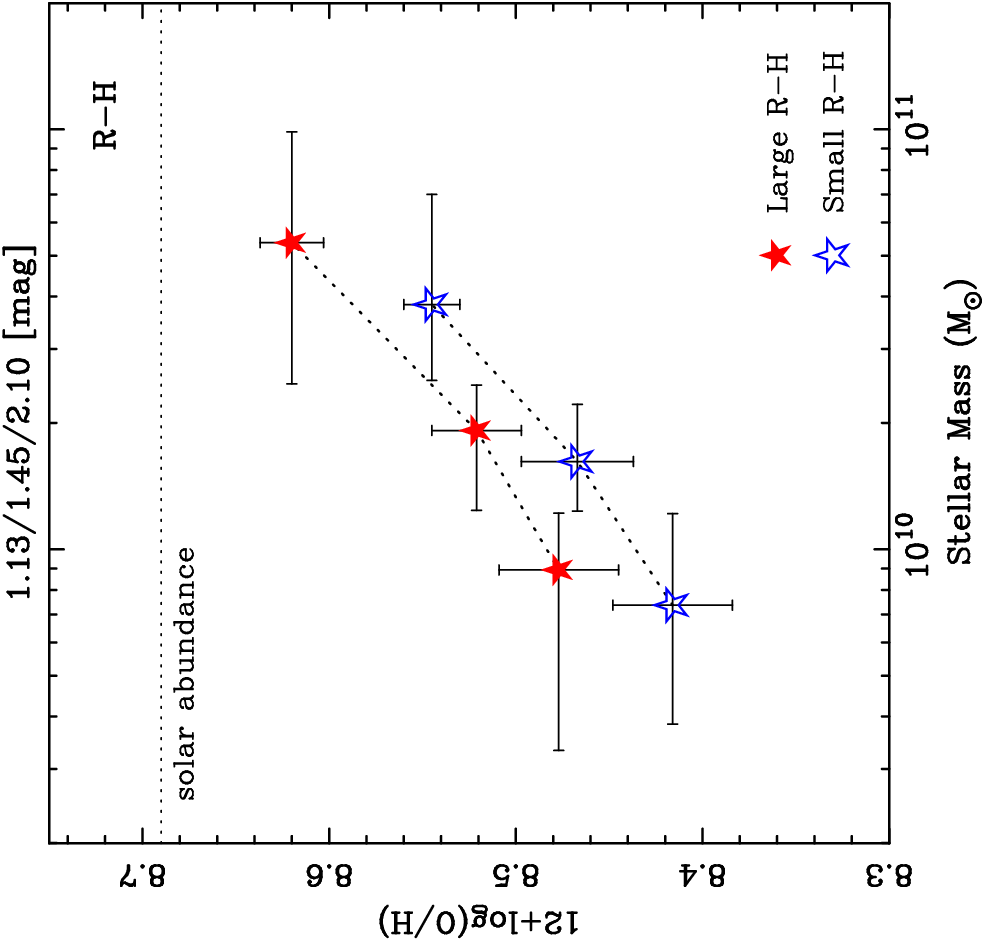}
\includegraphics[angle=-90,scale=0.33]{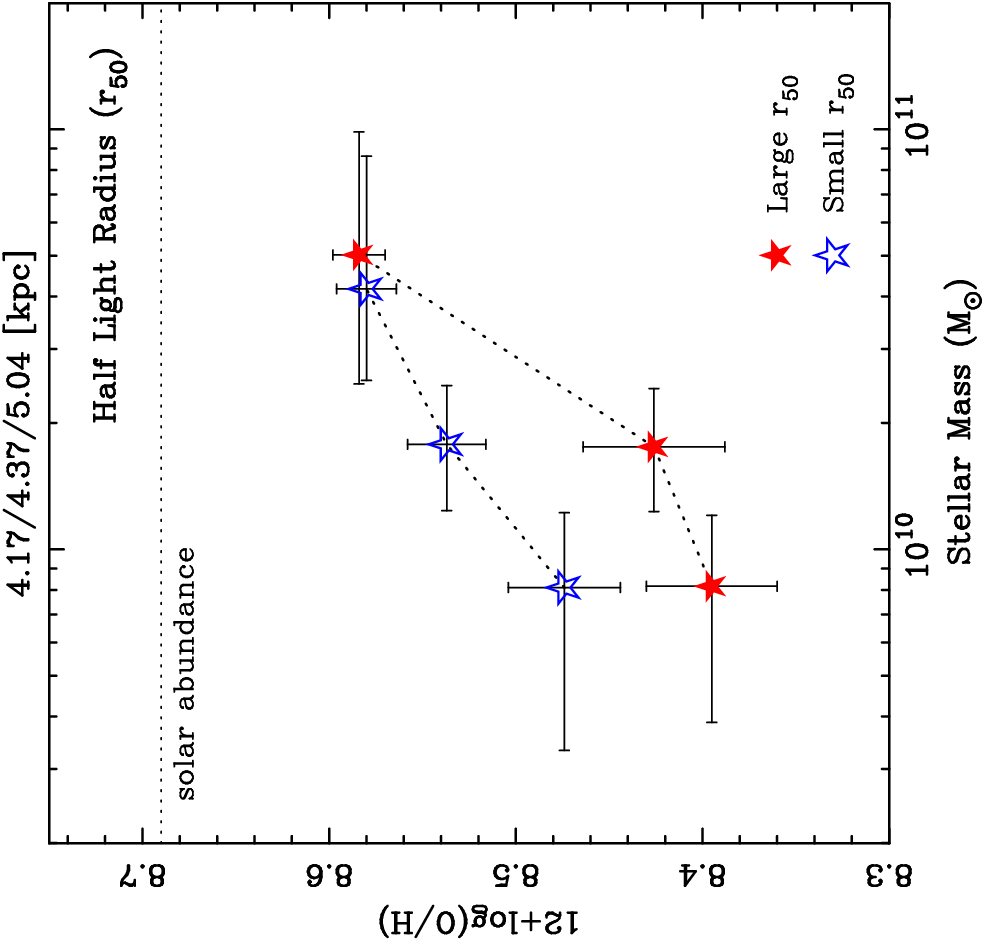}
\includegraphics[angle=-90,scale=0.33]{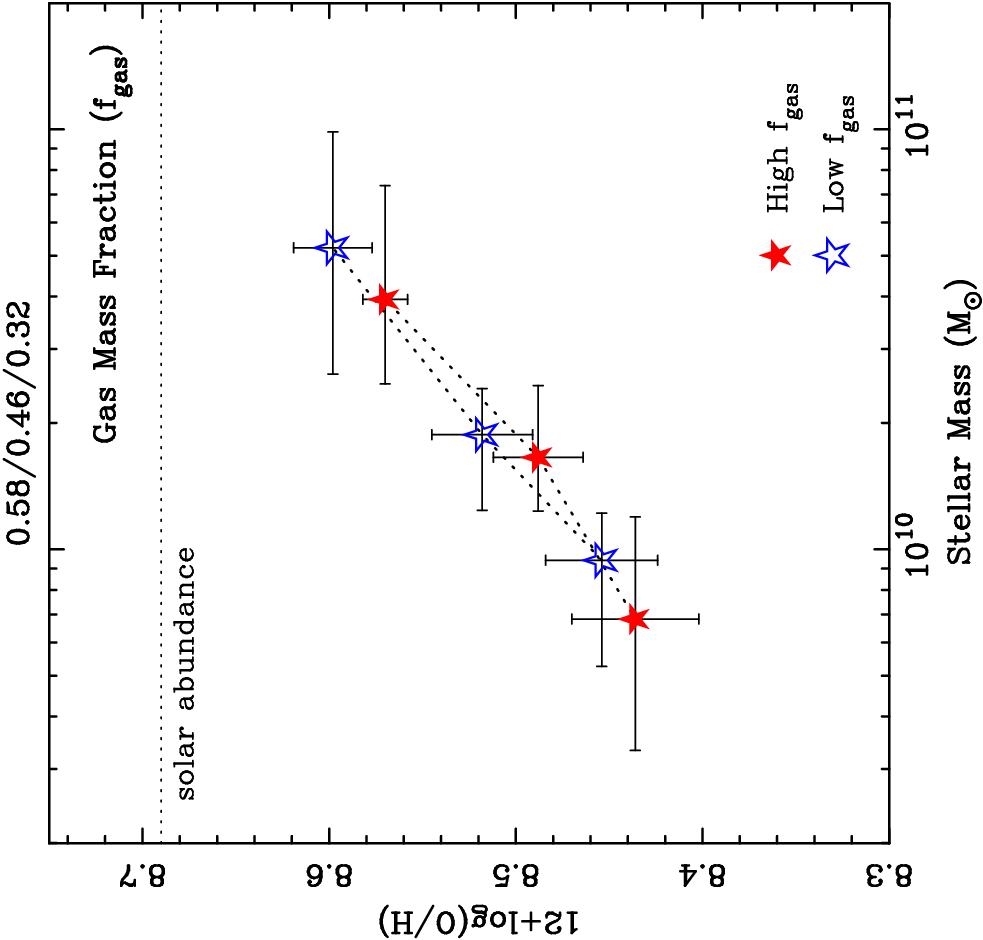}\\
\caption{The dependence of the mass-metallicity relation on the physical parameters. Here we present the dependence on intrinsic SFR from H$\alpha$ (top-left) and UV (top-middle) and specific SFR from H$\alpha$ (top-right) and UV (middle-left); in each case dust extinction correction is applied. We also present the dependence on colour excess $E(B-V)$ (middle-middle), observed $B-R$ colour (middle-right), observed $R-H$ colour (bottom-left), half light radius (bottom-middle), and gas mass fraction (bottom-right). In each stellar mass bin, we divide the sample into sub-sample by the parameter and the metallicity is derived from the stacking analysis. The error bars are derived by using the bootstrap method. The details are described in Section \ref{sec:MZR_dependence}. The results are also summarized in Table \ref{Tab3}. \label{fig:MZR_dependence}}
\end{figure*}

\subsection{Dependence of the Mass-Metallicity Relation on Galaxy Physical Parameters\label{sec:MZR_dependence}}
As presented in Figure \ref{fig:MZR}, the mass-metallicity relation at $z\sim1.4$ has a large scatter in metallicity. The observed scatter for our sample is calculated in the following way: The standard deviation of metallicities from the stacking data point is calculated in each stellar mass bin, where metallicity of objects with the [N\,{\sc ii}] non-detection (S/N$<$3.0) is fixed to the value corresponding to S/N=3.0 (method A). The resultant scatter ranges from 0.10 to 0.14 dex with a mean of 0.12 dex. The scatters calculated in this way, however, should be lower limits in the strict sense since the metallicities of many objects are upper limits. We presume the observed scatter in the statistical way with data including censored data based on the Kaplan-Meier (KM) estimator (method B). We use the Astronomy SURVival (\textsc{ASURV}) analysis package developed by \citet{Feigelson:1985p25911}. The resultant scatter is $\sim1.5$ times larger than that obtained by fixing the upper limits, ranging from 0.13 to 0.26 dex with a mean of 0.18 dex. The observed scatters in both ways are generally larger than the typical observational error of the [N\,{\sc ii}] detected objects. The intrinsic scatter can be estimated by subtracting the median value of observational errors from the observed scatter, resulting $0.07-0.13$ dex with a mean of 0.10 dex for method A and $0.12-0.25$ dex with a mean of 0.17 dex for method B. Since the estimation of the proper observational errors, however, is very difficult due to various systematic effects, we note that the intrinsic scatters may be upper limits even if we use method B.

Here, we examine the possible origin of the scatter, i.e., the dependence of the mass-metallicity relation on other parameters: The parameters we examined are the intrinsic SFR, sSFR, half light radius ($r_{50}$), $E(B-V)$, observed $B-R$ colour, and observed $R-H$ colour. The intrinsic SFR and sSFR are derived from both H$\alpha$ and the rest-frame UV luminosity density by using the relation by \citet{Kennicutt:1998p7465} and are corrected for the dust extinction and also the aperture effect as mentioned in Section \ref{sec:SpectralFitting}. The obtained SFR ranges from 8 M$_{\sun}$yr$^{-1}$ to 900 M$_{\sun}$yr$^{-1}$ with a median of 65 M$_{\sun}$yr$^{-1}$ from H$\alpha$, and 10 M$_{\sun}$yr$^{-1}$ to 3700 M$_{\sun}$yr$^{-1}$ with a median of 49 M$_{\sun}$yr$^{-1}$ from UV. The H$\alpha$ SFR agrees with the UV SFR within $\sigma\sim0.3$ dex, with no significant systematics. The sSFR derived from H$\alpha$ and UV range from 0.4 Gyr$^{-1}$ to 140 Gyr$^{-1}$ with a median of 3.3 Gyr$^{-1}$ (from H$\alpha$), and 0.6 Gyr$^{-1}$ to 30 Gyr$^{-1}$ with a median of 2.8 Gyr$^{-1}$ (from UV). The $r_{50}$, which is determined from the WFCAM $K-$band image by deconvolving the typical PSF size, ranges from 2.6 kpc to 7.3 kpc with a median of 4.5 kpc, and the $E(B-V)$, which is derived from the rest-frame UV colour, ranges from 0.02 mag to 0.94 mag with a median of 0.18. The observed $B-R$ and $R-H$ colours range from 0.04 mag to 2.00 mag with a median of 0.34, and 0.48 mag to 3.92 mag with a median of 1.49, respectively. The gas mass including H\,{\sc i} and H$_{2}$ is estimated from the SFR surface density computed from H$\alpha$ luminosity and size of the galaxy by assuming Kennicutt-Schmidt law \citep[K-S law; ][]{Kennicutt:1998p7470} with an index of $n=1.4$. We take the $r_{50}$ measured from K-band image, which traces $\sim$9000 \textrm{\AA} in the rest-frame, de-convolved by the seeing size as the intrinsic size of the galaxy. This indirect method is also used in the previous studies by \citet{Tremonti:2004p4119}  and \citet{Erb:2006p4143}. Although, the size of the region from which H$\alpha$ emission comes from H\,{\sc ii} region may be different from that of the stellar component, we assume that both have the same size. The gas mass fraction ($\equiv M_{\textrm{gas}} / (M_{\textrm{gas}}+M_{*})$) of our sample widely ranges from 0.1 to 0.9 with a median value of 0.45.

Our sample is divided into three stellar mass bins, i.e., mass bin 1 (9.500 $\le$ log($M_{*}/M_{\sun}$) $<$ 10.090), mass bin 2 (10.090 $\le$ log($M_{*}/M_{\sun}$) $<$ 10.391), and mass bin 3 (10.391 $\le$ log($M_{*}/M_{\sun}$) $\le$ 11.000). The sample in each stellar mass bin is then divided into two groups according to the median value of the parameters listed above. For each group, the stacking analysis is applied by using \textit{method 1} described in Section \ref{sec:SpectralFitting}. The resultant metallicity in each stellar mass bin and the parameter group is plotted as filled and open stars in Figure \ref{fig:MZR_dependence}, where the threshold of each parameter of the sample division is presented at the head of each panel. The threshold, the median value of each group, and the resultant metallicity are also summarized in Table \ref{Tab3}.

\textit{SFR and sSFR}: In Figure \ref{fig:MZR_dependence}, no clear dependence of the mass-metallicity relation on the SFR both from H$\alpha$ (top-left) and UV (top-right) can be seen. Although the higher SFR tends to show lower metallicity at the stellar mass bin 2, the metallicity does not strongly depend on the SFR at more massive (mass bin 3) and less massive (mass bin 1) bins. In the top-right and middle-left panels of Figure \ref{fig:MZR_dependence}, there are no clear dependence on sSFR. The absence of the clear dependence on the SFR implies that our sample selection by the expected H$\alpha$ flux does not affect the resulting mass-metallicity relation largely. In the local universe, the existence of the SFR \citep{Mannucci:2010p8026} and specific SFR \citep{Ellison:2008p7997} dependence on the mass-metallicity relation is reported. The sample at $z\sim0.1$ by \citet{Mannucci:2010p8026} covers the SFR range of $0.05$ to $10$ M$_{\sun}$yr$^{-1}$, while our sample covers that of $10$ to $200$ M$_{\sun}$yr$^{-1}$. However, the mean SFRs in each stellar mass bin of our sample only differ by a factor of 2. By extrapolating the results by \citet{Mannucci:2010p8026} toward the higher SFR range, the expected difference of the metallicity in the SFR range of our sample is $\sim0.02$ dex, which is comparable to or smaller than the bootstrap errors presented in Figure \ref{fig:MZR_dependence}, and thus would not be detectable. We may not be able to see clear dependence of the mass-metallicity relation on the SFR and sSFR partly because there is a large observational error and the parameter range of our high redshift sample is narrow. The clearer dependence may be found if we see galaxies with the lower level of SFR \citep[c.f., see ][]{Stott:2013p27513}. It is also worth noting that the dependence on SFR and sSFR at the massive end is small in the local sample by \citet{Mannucci:2010p8026} and \citet{Ellison:2008p7997}; especially the local dependence on sSFR is only visible at the stellar mass of $\la10^{10}$ M$_{\odot}$. As shown in Figure \ref{fig:MZR_zevol}, the redshift evolution can be seen in the low mass part. The redshift distribution in each SFR and sSFR groups, however, does not differ significantly, and thus no clear trend of the SFR and sSFR in the low mass part could not be due to the selection effect regarding to redshift.

\begin{figure}
\begin{center}
\includegraphics[scale=0.50,angle=270]{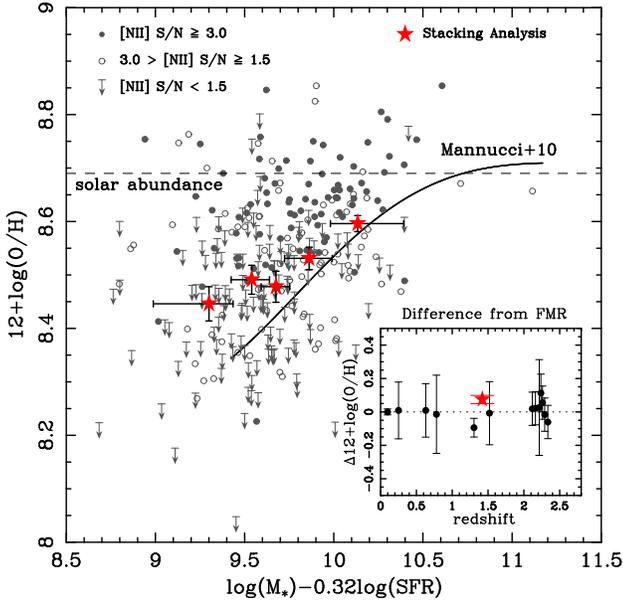}
\caption{The metallicity distribution of our sample against $\textrm{log}(\textrm{M}_{*})-0.32\textrm{log}(\textrm{SFR})$. Symbols of individual data points are the same as those in Figure \ref{fig:MZR}. The solid line shows the fundamental metallicity relation (FMR) proposed by \citet{Mannucci:2010p8026}. Both the stellar mass and metallicity of the FMR are converted so that the IMF and metallicity calibration are consistent with those we adopted. The bottom-right inset panel shows the difference from the FMR as a function of redshift with previous results taken from \citet{Mannucci:2010p8026} and \citet{Cresci:2011p20053}. The error bar of our sample is derived from the average error from the stacked spectra with the bootstrap resampling.\label{fig:FMR}}
\end{center}
\end{figure}

As we mentioned above, since there exists no clear dependence of the mass-metallicity relation on the intrinsic SFR at $z\sim1.4$, there seems to be no clear plane in the 3D-space with stellar mass, metallicity, and SFR, as reported in the FMR at $z\sim0.1$. In Figure \ref{fig:FMR}, the metallicities of our sample are plotted against log($M_{*}$)-$\alpha$log(SFR) with the average metallicity from the stacking analysis, where $\alpha$ is a projection parameter. At $z\sim0.1$, \citet{Mannucci:2010p8026} adopted the parameter of $\alpha=0.32$ that minimizes the residual scatter of median metallicity around the FMR. Here, we also assume the same projection parameter of $\alpha=0.32$. At $z\sim1.4$, no tight relation such as the FMR at $z\sim0.1$ by \citet{Mannucci:2010p8026} can be seen. The scatter of the metallicity against log($M_{*}$)-$\alpha$log(SFR) is also calculated by using the same method as that we described in Section log($M_{*}$)-$\alpha$log(SFR). If we fixed the metallicity of the non-detected objects to the upper limit values, the observed scatter is $\sim0.12$ dex. If we take into consideration the upper limits by using the KM estimator, the obtained scatter is $\sim0.17$ dex. Thus, the scatter of the mass-metallicity relation is not reduced significantly if the projection parameter changes, i.e., viewing the 3-D space from the different direction. As we mentioned above, again, this may be partly due to the large observational error and the narrower parameter range than the local SDSS sample. 

Although there is no clear surface at $z\sim1.4$ in the 3D-space, the overall position is close to the FMR at $z\sim0.1$. In Figure \ref{fig:FMR}, the result of the stacking analysis shows that, on average, our data points at $z\sim1.4$ are close to the FMR at $z\sim0.1$ by \citet{Mannucci:2010p8026}; our result, however, is shifted by $\ga0.1$ dex at the smaller log($M_{*}$)-0.32log(SFR) and thus lower metallicity part. Here we use the same Salpeter IMF and the N2 metallicity calibration for a fair comparison. The result does not change largely if we use the SFR derived form the UV luminosity density in the calculation of the log($M_{*}$)-0.32log(SFR). In the sub-panel of Figure \ref{fig:FMR}, the average difference from the local FMR as a function of redshift combining previous studies up to $z\sim2.5$ taken from \citet{Mannucci:2010p8026} and \citet{Cresci:2011p20053}. As we mentioned above, there is, however, a offset from the local FMR especially in the low mass part. We should note again that the sample at $z\sim0.1$ only covers the SFR range of $\la10$ M$_{\sun}$yr$^{-1}$, while our sample covers that of $\ga10$ M$_{\sun}$yr$^{-1}$.

\textit{Colour and colour excess}: Figure \ref{fig:MZR_dependence} shows that there exist weak dependences of $E(B-V)$ (middle-middle), observed $B-R$ (middle-right), and observed $R-H$ (bottom-left) colours on the mass-metallicity relation, where the observed $B-R$ and $R-H$ colours roughly correspond to $1900\textrm{\AA} - 2700\textrm{\AA}$ and $2700\textrm{\AA} - 6600\textrm{\AA}$ colours in the rest-frame, respectively. At any stellar mass range, objects with larger $E(B-V)$, $B-R$, and $R-H$ colours tend to show higher metallicity by $\sim0.05$ dex. The similarity of the dependence on the colour excess and the observed colours merely reflects the fact that the colour excess is derived from the rest-frame UV colour. Because there exists a correlation between these parameters and the stellar mass, i.e., massive galaxies tend to show larger $E(B-V)$, redder $B-R$ and $R-H$ colours. The dependence of the metallicity on each parameter in Figure \ref{fig:MZR_dependence} may be merely due to the mass-metallicity relation itself. If $E(B-V)$ was correlated with mass but did not influence the mass-metallicity relation, then the red and blue points in Figure \ref{fig:MZR_dependence} would all lie along a single line consistent with the mean mass-metallicity relation of our sample in Figure \ref{fig:MZR}. Therefore, the effect of the correlation to the stellar mass is likely to be small. Although the rest-frame coverage is slightly different, the colour dependence of the mass-metallicity relation is reported at $z\sim0.1$: \citet{Tremonti:2004p4119} find that galaxies with redder $g-i$ colour, which is k-corrected to $z=0.1$, tend to show higher metallicity with the difference of $\sim0.1$ dex at most. The average dependence on the colour excess for our sample is $\Delta$[12+log(O/H)]/$\Delta E(B-V)=0.56$ dex mag$^{-1}$, and the average colour dependences are $\Delta$[12+log(O/H)]/$\Delta (B-R)=0.26$ dex mag$^{-1}$ and $\Delta$[12+log(O/H)]/$\Delta (R-H)=0.16$ dex mag$^{-1}$, which are almost comparable to the local dependence by \citet{Tremonti:2004p4119}. It is interesting to note that the dependence on colour excess may be due to the fact that the dust-to-gas ratio depends on metallicity \citep[e.g.,][]{Galametz:2011p27349}. At fixed stellar mass and gas mass fraction, galaxies with higher metallicity would have higher dust attenuation and thus the larger colour excess. We should note, however, that the dust attenuation could also depend on the galaxy inclination, which should not show a correlation with metallicity.

\textit{Half light radius}: The bottom-middle panel of Figure \ref{fig:MZR_dependence} shows a relatively clear size dependence on the mass-metallicity relation; objects with the larger half light radius tend to show lower metallicity by up to $\sim0.11$ dex, except for the most massive part. The size dependence on the mass-metallicity relation at $z\sim1.4$ which is already reported by Y12 is confirmed with about 4 times larger sample. As Y12 pointed out, the aperture effect on metallicity is small for the typical galaxy size range of our sample unless the metallicity gradient changes drastically at high redshift. The dependence of metallicity on the half light radius of our sample is $\Delta$[12+log(O/H)]/$\Delta r_{50}=-0.08$ dex kpc$^{-1}$, which is consistent with Y12 and also agrees with the local dependence of $-0.1$ dex kpc$^{-1}$ on average presented by \citet{Ellison:2008p7997}. Our result that the size dependence is more prominent at lower mass also agrees with the local result by \citet{Ellison:2008p7997}.

\textit{Gas mass fraction}: In the bottom-right panel of Figure \ref{fig:MZR_dependence}, the dependence of the mass-metallicity relation on the gas mass fraction is presented; it is shown that galaxies with higher gas mass fraction tend to show lower metallicity in each stellar mass bin, though there exists the strong correlation between the gas mass fraction and the stellar mass (Yabe et al. 2013, in prep.), i.e., the dependence may be due to the mass-metallicity relation itself as we mentioned above. In fact, the data points in Figure \ref{fig:MZR_dependence} are mostly on the mass-metallicity relation itself. In the local universe, \citet{Bothwell:2013p25948} find that the mass-metallicity relation depends on the H\,{\sc i} gas mass and the similar scaling relation as the FMR found by \citet{Mannucci:2010p8026} in the parameter space of the stellar mass, metallicity, and H\,{\sc i} gas mass. On the other hand, the dependence on the H$_{2}$ gas mass, and the gas mass fraction is not clear, partly due to the small sample and the uncertainty of the CO-to-H$_{2}$ conversion factor.

\textit{Morphology}: In addition to these parameters, we also examine the morphology dependence of the mass-metallicity relation. A part of our sample galaxies is located in the \textit{CANDELS} \citep{Grogin:2011p17008,Koekemoer:2011p17029} region, where deep \textit{HST/ACS} and \textit{WFC3} imaging data are available. 50 objects are detected in the WFC3/F160W image, and 41 objects are detected in the both ACS/F814W and WFC3/F160W images. According to our eye-inspection, there seems to be a tendency that compact and bulge-dominated objects tend to be located on the upper side of the mass-metallicity relation, while diffuse and disc-dominated objects tend to reside in the lower side. We apply the CAS parameterization \citep[][]{Conselice:2003p21036} for the sample. The derived CAS-C (compactness) ranges from 2.0 to 3.5, CAS-A (asymmetry) ranges from 0.1 to 0.5, and CAS-S (clumpiness) ranges from 0.0 to 0.2. By using the CAS parameters, we divide the sample into two bins and stack the spectra. Although it is not clear, there seems to be a dependence of the mass-metallicity relation on the CAS parameter. Galaxies with higher CAS-C tend to show higher metallicity by $\sim$ 0.05 dex, while galaxies with higher CAS-A or CAS-S tend to show lower metallicity by $0.02-0.04$ dex. Since the size of the sample that can be examined for the morphology is still limited, however, the further discussions on the morphology dependence of the mass-metallicity relation will be presented in the future works.

One of the possible scenario that could explain the observed dependence of the mass-metallicity relation on various physical parameters is a presence of the galactic scale outflows \citep[see e.g.,][]{Dalcanton:2007p4660}. In Figure \ref{fig:MZR_Comparison_Models}, we show that the resultant mass-metallicity relation at $z\sim1.4$ is well explained by the theoretical models including moderately strong outflows. The presence of ubiquitous outflows in high redshift galaxies are reported in previous observations \citep{Weiner:2009p13953,Steidel:2010p10312,Newman:2012p23470}. The ejection of the enriched gas by outflows causes the decrease of the galaxy metallicity. \citet{Ellison:2008p7997} suggested that galaxies with smaller half light radii for a given stellar mass have more centrally concentrated stellar distribution. In the higher surface gravity, the evacuation of the enriched gas by the outflow would be inefficient. The size dependence of the mass-metallicity relation in Figure \ref{fig:MZR_dependence} could be explained by this effect. \citet{Ellison:2008p7997} also mentioned that higher sSFR leads to more efficient ejection of enriched gas with a downward shift in the mass-metallicity relation. Redder galaxies generally tend to show lower star-formation activity if the observed colour is directly related to the star-formation activity, and show inefficient outflow and metal ejection.

\section{Conclusions and Summary}
We present results from near-infrared spectroscopic observations of star-forming galaxies at $z\sim1.4$ with FMOS on the Subaru Telescope. We observed K-band selected galaxies at $1.2\le z_{ph} \le 1.6$ in the SXDS/UDS fields with $M_{*}\ge 10^{9.5} M_{\sun}$, and expected F(H$\alpha$) $\ge$ $5\times 10^{-17}$ erg s$^{-1}$ cm$^{-2}$. Among the observed $\sim1200$ targets, 343 objects show significant H$\alpha$ emission lines. The gas-phase metallicity is obtained from [N\,{\sc ii}]$\lambda$6584/H$\alpha$ line ratio, after excluding possible active galactic nuclei (AGNs). Due to the faintness of the [N\,{\sc ii}]$\lambda$6584 lines, we apply the stacking analysis and derive the mass-metallicity relation at $z\sim1.4$. We compare our results to previous results at different redshifts in the literature. The mass-metallicity relation at $z\sim1.4$ is located between those at $z\sim0.8$ and $z\sim2.2$; it is found that the metallicity increases with decreasing redshift from $z\sim3$ to $z\sim0$ at fixed stellar mass. Thanks to a large size of sample, we can study the dependence of the mass-metallicity relation on various galaxy physical properties. The dependence of the mass-metallicity relation on the SFR, which is previously found in the local universe, cannot be seen clearly in this work. No clear dependence on the sSFR can also be seen. We conclude that the difference from the local results may be partly due to the large observational errors and the narrow SFR range of our sample. Although our result shows no clear surface such as the fundamental metallicity relation (FMR) at $z\sim0.1$, by using the stacked spectra, we found that galaxies in our sample lie close to the local FMR in the higher metallicity part but an average of $\ga0.1$ dex higher in metallicity than the local FMR in the lower metallicity part. We also find trends that redder galaxies or galaxies with smaller half light radii show higher metallicity at fixed stellar mass. These observational facts partly can be explained by the scenario including evacuation of enriched gas by galactic-scale outflows. Since the degree of the parameter dependence of the mass-metallicity relation is very small compared to the typical statistical error in some cases, further observations in order to expand the sample size and the parameter range are desirable.

\section*{Acknowledgements}
We are grateful to the FMOS support astronomer Kentaro Aoki for his support during the observations. We also appreciate Soh Ikarashi, Kotaro Kohno, Kenta Matsuoka, and Tohru Nagao sharing fibres in their FMOS observations. We also thank the referee for insightful comments and suggestions which improved this paper. KY is financially supported by a Research Fellowship of the Japan Society for the Promotion of Science for Young Scientists. KO is supported by the Grant-in-Aid for Scientific Research (C) (24540230) from Japan Society for the Promotion of Science (JSPS). We acknowledge support for the FMOS instrument development from the UK Science and Technology Facilities Council (STFC). DB and ECL acknowledge support from STFC studentships. We would like to express our acknowledgement to the indigenous Hawaiian people for their understanding of the significant role of the summit of Mauna Kea in astronomical research.

\end{document}